\documentstyle[12pt]{article}

\topmargin - 20mm
\begin{document}

\title{Ising Model and $L$ -- Function}

\author{Yury M. Zinoviev\thanks{This work is supported in part by the
Russian Foundation for Basic Research (Grant No. 00 -- 01 -- 00083)} \\
Steklov Mathematical Institute, \\ Gubkin St. 8, Moscow 117966, GSP -- 1,
Russia \\ e -- mail: zinoviev@mi.ras.ru}

\date{}
\maketitle

\vskip 1cm

\noindent {\bf Abctract}. The correlation functions are calculated for
the two dimensional Ising model with free boundary conditions and 
the two dimensional Ising model with periodic boundary conditions.

\vskip 1cm

\section{Introduction}

\noindent Onsager and Kaufman \cite{1} -- \cite{3} invented the 
formula for the partition function of the two dimensional Ising model.
Another method for the calculation of the partition function
was proposed by Kac and Ward \cite{4}. They considered simultaneously two
formulae: the determinant of the special matrix $I + T$ ($I$ is the identity
matrix) is proportional to the partition function of the Ising model and it
is proportional to the square of the partition function. For the proof of
the first formula they used a topological statement. Sherman \cite{5}, 
\cite{6} constructed a counter -- example for this statement. Hurst and
Green \cite{7} proposed to use for the calculation of the Ising model
partition function not a determinant but a Pfaffian of some special matrix.
This method was improved by Kasteleyn \cite{8}, Fisher \cite{9}, McCoy and
Wu \cite{10}. McCoy and Wu \cite{10} obtained the wrong formula connecting 
the Pfaffians with the Ising model partition function. The proper formula 
of this type is obtained in the paper \cite{11}. Sherman \cite{5}, \cite{6}
gave some arguments for the equality
\begin{equation}
\label{1.1}
Z^{2} = C(\beta )\det (I + T)
\end{equation}
where $Z$ is the partition function of the two dimensional Ising model with
the free boundary conditions and $C(\beta )$ is the positive function of the
inverse temperature $\beta $. In the paper \cite{12} the following formula
\begin{equation}
\label{1.2}
Z^{2} = C(\beta )\det (I - T)
\end{equation}
is proved. For the rectangular lattice the expression (\ref{1.2}) is
independent of the sign of the matrix $T$. For an arbitrary lattice the
formula (\ref{1.1}) is wrong. If the matrices $T^{k}$ satisfy some estimate,
then
\begin{equation}
\label{1.3}
\det (I - T) = \exp \{ - \sum_{k = 1}^{\infty } k^{- 1}\hbox{tr} \, T^{k}\}.
\end{equation}
By the definition the partition function $Z > 0$. If the numbers
$\hbox{tr} \, T^{k}$ are real, then the equalities (\ref{1.2}) and 
(\ref{1.3}) imply
\begin{equation}
\label{1.4}
Z(C(\beta ))^{- 1/2} = 
\exp \{ - 1/2 \sum_{k = 1}^{\infty } k^{- 1}\hbox{tr} \, T^{k}\}.
\end{equation}
The bulk of Sherman papers \cite{5}, \cite{6} and the subsequent 
Burgoyne paper \cite{13} were devoted to the proof of Feynman conjecture.
Due to Feynman conjecture the Ising model partition function is 
proportional to some infinite formal product. The right hand side of the
equality (\ref{1.4}) is the product. By calculating the numbers
$\hbox{tr} \, T^{k}$ it is possible to show that the equality (\ref{1.4})
is a correct statement for Feynman conjecture.

Hashimoto studied some special infinite product. He called these products
zeta functions of finite graphs in the paper \cite{14} and $L$ -- functions
of finite graphs in the paper \cite{15}. It is possible to prove that the
right hand side of the equality (\ref{1.3}) is one over the Hashimoto
$L$ -- function (zeta function). The definition (\ref{1.3}) seems suitable
since the series in (\ref{1.3}) is convergent if the numbers 
$\hbox{tr} \, T^{k}$ satisfy some estimate.

The paper \cite{4} -- \cite{6}, \cite{12}, \cite{13} used the well -- known
van der Waerden formula \cite{16} for the Ising model partition function.
Similar formula for the correlation functions was obtained in the paper 
\cite{17}. By using these formulae and the formula (\ref{1.4}) we caculate 
the thermodynamic limit of the free energy and the correlation functions of
the two dimensional Ising model with free boundary conditions. Similar
results are obtained for periodic boundary conditions.

In the second section we discuss Hashimoto results \cite{14}, \cite{15}.
The third section is devoted to formula \cite{17} for the correlation 
functions. In the fourth section we study the two dimensional Ising model
with free boundary conditions. The fifth section is devoted to the two
dimensional Ising model with periodic boundary conditions.

\section{$L$ -- Function}
\setcounter{equation}{0}

\noindent Let $s$ be a complex number. Then for $\Re s > 1$ Riemann zeta
function
\begin{equation}
\label{2.1}
\zeta (s) = \sum_{n = 1}^{\infty} n^{- s}
\end{equation}
is an analytic function of the variable $s$. Euler showed that for $\Re s > 1$
(\cite{18},formula 17.7.2) 
\begin{equation}
\label{2.2}
\zeta (s) = \prod_{p} (1 - p^{- s})^{- 1}
\end{equation}
where the product is extended over the set of all prime numbers 
$p = 2,3,5,7,...$.

Let $n > 1$ be a fixed natural number and let $m$ be any natural number.
Let us consider the functions $\chi (m)$ such that

$\chi (m) = \chi (m^{\prime})$, if $m = m^{\prime} \, \hbox{mod} \, n$,

$\chi (1) = 1$,

$\chi (m) = 0$, if the greatest common divisor $(m,n)$ of the natural
numbers $m$ and $n$ is not one,

\begin{equation}
\label{2.3}
\chi (m)\chi (m^{\prime}) = \chi (mm^{\prime}).
\end{equation}
Such function $\chi (m)$ is called a modulo $n$ character.

Let $n > 1$ be a fixed natural number and let $\chi $ be a modulo $n$
character. Then the series
\begin{equation}
\label{2.4}
L(s,\chi ) = \sum_{m = 1}^{\infty} \chi (m)m^{- s}
\end{equation}
for $\Re s > 1$ is called the $L$ -- series. The $L$ -- series was
introduced by Dirichlet. The $L$ -- series and Riemann zeta function 
have many common properties. The Euler product (\ref{2.2}) analogue is 
(\cite{18}, formula 17.8.5)
\begin{equation}
\label{2.5}
L(s,\chi ) = \prod_{m = 1}^{\infty} (1 - \chi (p)p^{- s})^{- 1}
\end{equation}
where $\Re s > 1$ and the product is extended over the set of all prime 
numbers.

A function $\chi (m)$ of the natural number $m$ satisfying the condition
(\ref{2.3}) only is called a character. The function $\chi (m,s) = m^{- s}$
gives an example of a character. Let us consider the series (\ref{2.4})
with the character $\chi (m)$. When the character $\chi (m) \equiv 1$ the
series (\ref{2.4}) coincides with Riemann zeta function (\ref{2.1}). When the 
character $\chi (m)$ is a modulo $n$ character the series (\ref{2.4}) is
Dirichlet $L$ -- series.

Let us introduce $L$ -- function of the finite graph. Let $X$ be a 
finite graph and let ${\bf e}$ be an oriented edge of a graph $X$.
The oriented edge ${\bf e}$ is defined by the pair of the vertices of 
a graph $X$: the beginning $b({\bf e})$ and the end $f({\bf e})$ of
the oriented edge ${\bf e}$. A closed path is a sequence of the oriented
edges $C = ({\bf e}_{1},...,{\bf e}_{k})$ such that
\begin{equation}
\label{2.6}
b({\bf e}_{i + 1}) = f({\bf e}_{i}), \, i = 1,...,k - 1, \,
b({\bf e}_{1}) = f({\bf e}_{k}). 
\end{equation}
Let us denote $b(C) = b({\bf e}_{1})$ and $f(C) = f({\bf e}_{k})$. For a
closed path $b(C) = f(C)$. Let ${\bf e}^{- 1}$ be such oriented edge that
$b({\bf e}^{- 1}) = f({\bf e})$, $f({\bf e}^{- 1}) = b({\bf e})$. The
following pairs of the closed paths are regarded homotopic equivalent
\begin{equation}
\label{2.7}
({\bf e},{\bf e}^{- 1}) \sim C(b({\bf e}))
\end{equation}
where the path $C(b({\bf e}))$ consists of the only vertex $b({\bf e})$;
\begin{equation}
\label{2.8}
({\bf e}_{1},...,{\bf e}_{i - 1},{\bf e}_{i},{\bf e}_{i}^{- 1},
{\bf e}_{i + 2},...,{\bf e}_{k}) \sim
({\bf e}_{1},...,{\bf e}_{i - 1},{\bf e}_{i + 2},...,{\bf e}_{k})
\end{equation}
where $i = 2,...,k - 2$, $k > 3$;
\begin{equation}
\label{2.9}
({\bf e}_{1},{\bf e}_{1}^{- 1},{\bf e}_{3},...,{\bf e}_{k}) \sim
({\bf e}_{3},...,{\bf e}_{k});
\end{equation}
\begin{equation}
\label{2.10}
({\bf e}_{1},...,{\bf e}_{k - 2},{\bf e}_{k - 1},{\bf e}_{k - 1}^{- 1})
\sim ({\bf e}_{1},...,{\bf e}_{k - 2});
\end{equation}
\begin{equation}
\label{2.11}
({\bf e}_{1},{\bf e}_{2},...,{\bf e}_{k - 1},{\bf e}_{1}^{- 1}) \sim
({\bf e}_{2},...,{\bf e}_{k - 1}).
\end{equation}
Two closed paths are regarded homotopic if they are related by the 
equivalence relation generated by the relations (\ref{2.7}) -- (\ref{2.11}).

\noindent {\bf Lemma 2.1.} {\it The closed paths} 
$({\bf e}_{k},{\bf e}_{1},...,{\bf e}_{k - 1})$ {\it and}
$({\bf e}_{1},...,{\bf e}_{k})$ {\it are homotopic}.

\noindent {\it Proof}. The equivalence relation (\ref{2.11}) implies
\begin{equation}
\label{2.12}
({\bf e}_{k},{\bf e}_{1},...,{\bf e}_{k},{\bf e}_{k}^{- 1}) \sim
({\bf e}_{1},...,{\bf e}_{k}).
\end{equation}
The equivalence relation (\ref{2.10}) implies
\begin{equation}
\label{2.13}
({\bf e}_{k},{\bf e}_{1},...,{\bf e}_{k},{\bf e}_{k}^{- 1}) \sim
({\bf e}_{k},{\bf e}_{1},...,{\bf e}_{k - 1}).
\end{equation}
It follows from the equivalence relations (\ref{2.12}) and (\ref{2.13})
that the closed paths $({\bf e}_{1},...,{\bf e}_{k})$ and
$({\bf e}_{k},{\bf e}_{1},...,{\bf e}_{k - 1})$ are homotopic. The lemma 
is proved.

The equivalence relations (\ref{2.7}) -- (\ref{2.11}) imply that every
closed path is homotopic to a path consisting of the only vertex or is
homotopic to a reduced closed path $({\bf e}_{1},...,{\bf e}_{k})$:
\begin{equation}
\label{2.14}
b({\bf e}_{i + 1}) = f({\bf e}_{i}), \,
f({\bf e}_{i + 1}) \neq b({\bf e}_{i}), \, i = 1,...,k - 1, \,
b({\bf e}_{1}) = f({\bf e}_{k}), \, f({\bf e}_{1}) \neq b({\bf e}_{k}).
\end{equation}
In view of Lemma 2.1 $k$ reduced closed paths 
$({\bf e}_{1},...,{\bf e}_{k})$, 
$({\bf e}_{k},{\bf e}_{1},...,{\bf e}_{k - 1})$,...,
$({\bf e}_{2},...,{\bf e}_{k},{\bf e}_{1})$ are homotopic to each other.
They define the equivalence class called the oriented reduced cycle. We
denote it by $[({\bf e}_{1},..., {\bf e}_{k})]$. An oriented reduced cycle,
or a reduced closed path $({\bf e}_{1},...,{\bf e}_{k})$ representing it,
is called non -- primitive, if there exists a positive integer $l$
($1 \leq l < k$) such that $({\bf e}_{1},...,{\bf e}_{k}) =
({\bf e}_{l + 1},...,{\bf e}_{k},{\bf e}_{1},...,{\bf e}_{l})$; and
otherwise it is called primitive.

If two closed paths $C_{1} = ({\bf e}_{1},...,{\bf e}_{k})$,
$C_{2} = ({\bf e}_{k + 1},...,{\bf e}_{k + l})$ have the same beginning
vertex: $b({\bf e}_{1}) = b({\bf e}_{k + 1})$, it is possible to define the
product $C_{1}\cdot C_{2} = 
({\bf e}_{1},...,{\bf e}_{k},{\bf e}_{k + 1},...,{\bf e}_{k + l})$. If
$b({\bf e}_{1}) = b({\bf e}_{k + 1})$ and ${\bf e}_{k + 1} \neq {\bf e}_{1}$,
then the product $C_{1}\cdot C_{3} = 
({\bf e}_{1},...,{\bf e}_{k},{\bf e}_{k + 1},...,{\bf e}_{1}^{- 1})$
of two reduced closed paths $C_{1} = ({\bf e}_{1},...,{\bf e}_{k})$ and
$C_{3} = ({\bf e}_{k + 1},..,{\bf e}_{1}^{- 1})$ is not a reduced closed
path. The product $C_{1}\cdot C_{2} = ({\bf e}_{1},...,{\bf e}_{k + l})$
of two reduced closed paths $C_{1} = ({\bf e}_{1},...,{\bf e}_{k})$ and
$C_{2} = ({\bf e}_{k + 1},...,{\bf e}_{k + l})$ with the same beginning
edge: ${\bf e}_{1} = {\bf e}_{k + 1}$ is the reduced closed path.

Let a unitary matrix $\rho (C)$ in $n$ dimensional space correspond with any 
reduced closed path $C$ and satisfy the following conditions: if two reduced 
closed paths $C_{1} = ({\bf e}_{1},...,{\bf e}_{k})$ and
$C_{2} = ({\bf e}_{k + 1},...,{\bf e}_{k + l})$ have the same beginning
edge: ${\bf e}_{1} = {\bf e}_{k + 1}$ then 
\begin{equation}
\label{2.15}
\rho (C_{1}\cdot C_{2}) = \rho (C_{1}) \rho (C_{2});
\end{equation}
there exists a unitary matrix $\gamma $ for any reduced closed path 
$({\bf e}_{1},...,{\bf e}_{k})$ such that
\begin{equation}
\label{2.16}
\rho (({\bf e}_{k},{\bf e}_{1},...,{\bf e}_{k - 1})) =
\gamma \rho (({\bf e}_{1},...,{\bf e}_{k})) \gamma^{- 1}.
\end{equation}
{\bf Lemma 2.2.} {\it Let a unitary matrix} 
$\rho ({\bf e}_{1};{\bf e}_{2})$ {\it correspond with any pair}
${\bf e}_{1},{\bf e}_{2}$ {\it of the oriented edges of the graph}
$X$ {\it such that} $f({\bf e}_{1}) = b({\bf e}_{2})$,
$b({\bf e}_{1}) \neq f({\bf e}_{2})$. {\it For any reduced closed path}
$({\bf e}_{1},...,{\bf e}_{k})$ {\it we define the unitary matrix}
\begin{equation}
\label{2.17}
\rho (({\bf e}_{1},...,{\bf e}_{k})) = \rho ({\bf e}_{1};{\bf e}_{2})
\rho ({\bf e}_{2};{\bf e}_{3}) \cdots \rho ({\bf e}_{k - 1};{\bf e}_{k})
\rho ({\bf e}_{k};{\bf e}_{1}).
\end{equation}
{\it Then the matrix} (\ref{2.17}) {\it satisfies the relations}
(\ref{2.15}) {\it and} (\ref{2.16}).

\noindent {\it Proof.} Let two reduced closed paths 
$C_{1} = ({\bf e}_{1},...,{\bf e}_{k})$ and
$C_{2} = ({\bf e}_{k + 1},...,{\bf e}_{k + l})$ have the same beginning
edge: ${\bf e}_{1} = {\bf e}_{k + 1}$. Then the definition (\ref{2.17})
implies the relation (\ref{2.15}).

Let $({\bf e}_{1},...,{\bf e}_{k})$ be a reduced closed path. Then the
definition (\ref{2.17}) implies the relation (\ref{2.16}) with the unitary 
matrix $\gamma = \rho ({\bf e}_{k};{\bf e}_{1})$. The lemma is proved.

By a labelling on the set of non -- oriented edges of the graph $X$ we
mean an assignment $e \rightarrow u(e) = u({\bf e}) = u({\bf e}^{- 1})$,
where $u(e)$ are idependent variables for the different non -- oriented
edges. We denote them simply by ${\bf u} = \{ u(e)\} $. We put
\begin{equation}
\label{2.18}
{\bf u}^{C} = \prod_{i = 1}^{k} u({\bf e}_{i})
\end{equation}
where $C = ({\bf e}_{1},...,{\bf e}_{k})$ is a reduced closed path. The
$L$ -- function of $X$ attached to $\rho $ is defined by the following 
formal infinite product similar to the product (\ref{2.5})
\begin{equation}
\label{2.19}
L({\bf u},\rho ;X) = \prod_{[C]: \, primitive} 
\det (I_{n} - \rho (C){\bf u}^{C})^{- 1}
\end{equation}
where the product is extended over the set of primitive oriented reduced
cycles $[C]$ and $I_{n}$ is the idetity matrix in the $n$ dimensional space.

If the graph $X$ is connected, then for any reduced closed path $C$ there
exists a homotopic closed path $\langle C\rangle $ such that 
$b(\langle C\rangle ) = p$ where $p$ is the fixed vertex. In view of the 
relation (\ref{2.9}) we may consider that the beginning edge of the closed 
path $\langle C \rangle $ is the fixed oriented edge ${\bf e}$ and 
$b({\bf e}) = p$. Hashimoto \cite{14}, \cite{15} considered the unitary 
representation $\rho $ of the group of the classes of homotopically equivalent 
closed paths $\langle C \rangle $ with the fixed initial vertex $p$. 
If we substitute the matrix $\rho (\langle C\rangle )$ instead of the matrix
$\rho (C)$ in the definition (\ref{2.19}), we obtain Hashimoto definition
\cite{15}. In the paper \cite{14} the function $L({\bf u},\rho ;X)$ is
denoted by $Z_{X}({\bf u};\rho )$ and it is called a zeta function.

The definition (\ref{2.19}) is formal. We transform it into another form.
We denote by $\alpha_{i} (C), i = 1,...,n$, the eigenvalues of the unitary
matrix $\rho (C)$. Since the matrix $\rho (C)$ is unitary, 
\begin{equation}
\label{2.20}
|\alpha_{i} (C)| = 1, \, i = 1,...,n. 
\end{equation} 
Taking the logarithm of (\ref{2.19}) we have
\begin{equation}
\label{2.21}
\ln L({\bf u},\rho ;X) = \sum_{[C]: \, primitive}
\ln [\det (I_{n} - \rho (C){\bf u}^{C})^{- 1}]
\end{equation}
\begin{eqnarray}
\label{2.22}
\ln [\det (I_{n} - \rho (C){\bf u}^{C})^{- 1}] =
- \sum_{i = 1}^{n} \ln (1 - \alpha_{i} (C){\bf u}^{C}) = \nonumber \\
\sum_{i = 1}^{n} \sum_{k = 1}^{\infty } k^{- 1}(\alpha_{i} (C))^{k}
({\bf u}^{C})^{k} =
\sum_{k = 1}^{\infty } k^{- 1}(\hbox{tr} \, \rho (C)^{k})
({\bf u}^{C})^{k}.
\end{eqnarray}
The substiturion of the equality (\ref{2.22}) into the equality (\ref{2.21})
gives
\begin{equation}
\label{2.23}
\ln L({\bf u},\rho ;X) = \sum_{[C]: \, primitive} \sum_{k = 1}^{\infty }
k^{- 1}({\bf u}^{C})^{k}\hbox{tr} \, \rho (C)^{k}.
\end{equation}
Any primitive oriented reduced cycle $[({\bf e}_{1},...,{\bf e}_{l})]$
consists of $l$ different reduced closed paths: 
$({\bf e}_{1},...,{\bf e}_{l})$, 
$({\bf e}_{l},{\bf e}_{1},...{\bf e}_{l - 1})$,...,
$({\bf e}_{2},...,{\bf e}_{l},{\bf e}_{1})$. Due to the definition 
(\ref{2.18})
\begin{equation}
\label{2.24}
{\bf u}^{(e_{l},e_{1},...,e_{l - 1})} = 
{\bf u}^{(e_{1},...,e_{k})}.
\end{equation}
for the reduced closed paths $({\bf e}_{l},{\bf e}_{1},...,{\bf e}_{l - 1})$ 
and $({\bf e}_{1},...,{\bf e}_{l})$. It follows from the equalities 
(\ref{2.16}) and (\ref{2.24}) that the equality (\ref{2.23}) may be 
rewritten as
\begin{equation}
\label{2.25}
\ln L({\bf u},\rho ;X) = \sum_{C: \, primitive} \sum_{k = 1}^{\infty }
(k|C|)^{- 1}({\bf u}^{C})^{k}\hbox{tr} \, \rho (C)^{k}
\end{equation}
where $C$ runs over the set of primitive reduced closed paths
$C = ({\bf e}_{1},...,{\bf e}_{l})$ and the length $|C| = l$.

The definition (\ref{2.18}) implies
\begin{equation}
\label{2.26}
({\bf u}^{C})^{k} = {\bf u}^{C^{\times k}}.
\end{equation}
The equalities (\ref{2.15}), (\ref{2.25}), (\ref{2.26}) and the equality
$|C^{\times k}| = k|C|$ imply
\begin{equation}
\label{2.27}
\ln L({\bf u},\rho ;X) = \sum_{C: \, primitive} \sum_{k = 1}^{\infty }
|C^{\times k}|^{- 1}{\bf u}^{C^{\times k}} \hbox{tr} \, \rho (C^{\times k}).
\end{equation}
Any reduced closed path has the form $C^{\times k}$ where $k$ is a natural
number and $C$ is a primitive reduced closed path. Then the equality 
(\ref{2.27}) may be rewritten as
\begin{equation}
\label{2.28}
\ln L({\bf u},\rho ;X) = \sum_{C \in RC(X)} |C|^{- 1}{\bf u}^{C}
\hbox{tr} \, \rho (C)
\end{equation}
where $C$ runs over the set $RC(X)$ of all reduced closed paths on 
the graph $X$.

The number of the non -- oriented edges which is incident to a vertex
$p$ is called  $v(p)$, the valency of $p$. Let
\begin{equation}
\label{2.29}
v = \max_{p} v(p).
\end{equation}
Let us construct a reduced closed path with the initial vertex $p$.
For the first edge we have $v(p)$ possibilities. For any other edge
the number of possibilities is less than $v - 1$. Thus the total
number of reduced closed paths of the length $l$ is less than 
$\# (VX)v(v - 1)^{l - 1}$ where $\# (VX)$ is the total number of vertices 
of the graph $X$. The relations (\ref{2.20}) imply that the series
(\ref{2.28}) is absolutely convergent if the following estimate is valid
\begin{equation}
\label{2.30}
\max_{e} |u({\bf e})| < (v - 1)^{- 1}.
\end{equation}
{\bf Definition.} {\it The} $L$ -- {\it function} {\it of the finite graph}
$X$ {attached to} $\rho $ {\it defined by}
\begin{equation}
\label{2.31}
L({\bf u},\rho ;X) = \exp \{ \sum_{C \in RC(X)} |C|^{- 1}
{\bf u}^{C} \hbox{tr} \, \rho (C) \}
\end{equation}
{\it where} $C$ {\it runs over the set} $RC(X)$ {\it of all reduced
closed paths on the graph} $X$.

Let $\# ({\bf E}X)$ be the total number of oriented edges of the graph $X$.

\noindent {\bf Theorem 2.3.} {\it Let unitary matrix}
$\rho ({\bf e}_{1};{\bf e}_{2}) = 
\{ (\rho ({\bf e}_{1};{\bf e}_{2}))_{k_{1}k_{2}}, k_{1},k_{2} = 1,...,n\} $
{\it correspond with any pair} ${\bf e}_{1},{\bf e}_{2}$
{\it of oriented edges of the graph} $X$, {\it such that} 
$f({\bf e}_{1}) = b({\bf e}_{2})$, $b({\bf e}_{1}) \neq f({\bf e}_{2})$.
{\it Let us define} $(n\# ({\bf E}X))\times (n\# ({\bf E}X))$ -- {\it matrix}
\begin{equation}
\label{2.32}
T({\bf u},\rho )_{(k_{1},e_{1}),(k_{2},e_{2})} 
= \left\{ {u({\bf e}_{1})(\rho ({\bf e}_{1};{\bf e}_{2}))_{k_{1}k_{2}}, 
\hskip 1cm f({\bf e}_{1}) = b({\bf e}_{2}), \, 
b({\bf e}_{1}) \neq f({\bf e}_{2}),} \atop 
{0, \hskip 1cm  otherwise.} \right.
\end{equation}
{\it If the estimate} (\ref{2.30}) {\it is fulfilled, then}
\begin{equation}
\label{2.33}
L({\bf u},\rho ;X) = \det (I - T({\bf u},\rho ))^{- 1}
\end{equation}
{\it where} $L$ -- {\it fuction} $L({\bf u},\rho ;X)$ {\it is defined by
the equality} (\ref{2.31}).

\noindent {\it Proof.} Analogously to the equality (\ref{2.22}) we get
\begin{equation}
\label{2.34}
\ln [\det (I - T({\bf u},\rho ))^{- 1}] =
\sum_{k = 1}^{\infty } k^{- 1}\hbox{tr} \, T({\bf u},\rho )^{k}.
\end{equation}
Due to the definitions (\ref{2.17}), (\ref{2.18}) and (\ref{2.32}) we have
\begin{equation}
\label{2.35}
\hbox{tr} \, T({\bf u},\rho )^{k} = 
\sum_{{C \in RC(X),} \atop {|C| = k}} {\bf u}^{C} 
\hbox{tr} \, \rho (C)
\end{equation}
where $C$ runs over the set of all reduced closed paths of the length $k$.
The substitution of the equality (\ref{2.35}) into the right hand side of
the equality (\ref{2.34}) gives
\begin{equation}
\label{2.36}
\ln [\det (I - T({\bf u},\rho ))^{- 1}] =
\sum_{C \in RC(X)} |C|^{- 1}{\bf u}^{C} 
\hbox{tr} \, \rho (C)
\end{equation}
where $C$ runs over the set $RC(X)$ of all reduced closed paths on
the graph $X$. The equalities (\ref{2.31}) and (\ref{2.36}) imply the 
equality (\ref{2.33}). The theorem is proved.

\section{Ising Model}
\setcounter{equation}{0}

\noindent We consider a rectangular lattice on the plane formed by
the points with integral Cartesian coordinates $x = k_{1}$, $y = k_{2}$,
$M_{1}^{\prime } \leq k_{1} \leq M_{1}$, 
$M_{2}^{\prime } \leq k_{2} \leq M_{2}$, and the corresponding 
horizontal and vertical edges connecting these vertices. We denote
this graph by $G(M_{1}^{\prime },M_{2}^{\prime };M_{1},M_{2})$.
Let a graph $G$ be embedded in a rectangular lattice on the plane.
Let all the vertices from the boundaries of all the edges of a graph
$G$ be included into the set of the vertices of a graph $G$. The 
cell complex $P(G)$ is called the set consisting of the cells (vertices,
edges, faces). A vertex of $P(G)$ is called a cell of dimension $0$.
It is denoted by $s_{i}^{0}$. An edge of $P(G)$ is called a cell of
dimension $1$. It is denoted by $s_{i}^{1}$. A face of $P(G)$ is
called a cell of dimension $2$. It is denoted by $s_{i}^{2}$. We
suppose that $P(G)$ contains all the faces whose all boundary edges
are included into a graph $G$. We denote by ${\bf Z}_{2}^{add}$ the
group of modulo $2$ residuals. The modulo $2$ residuals are multiplied 
by each other and the group ${\bf Z}_{2}^{add}$ is a field. To every
pair of the cells $s_{i}^{p}$, $s_{j}^{p - 1}$ there corresponds the
number $(s_{i}^{p}:s_{j}^{p - 1}) \in {\bf Z}_{2}^{add}$ (incidence
number). If the cell $s_{j}^{p - 1}$ is included into the boundary of
the cell $s_{i}^{p}$, then the incidence number 
$(s_{i}^{p}:s_{j}^{p - 1}) = 1$. Otherwise the incidence number
$(s_{i}^{p}:s_{j}^{p - 1}) = 0$. For any pair of the cells $s_{i}^{2}$,
$s_{j}^{0}$ the incidence numbers satisfy the condition
\begin{equation}
\label{3.1}
\sum_{m} (s_{i}^{2}:s_{m}^{1})(s_{m}^{1}:s_{j}^{0}) = 0 \,
\hbox{mod} \, 2.
\end{equation}
Indeed, if the vertex $s_{j}^{0}$ is not included into the boundary of
the square $s_{i}^{2}$, then the condition (\ref{3.1}) is fulfilled.
If the vertex $s_{j}^{0}$ is included into the boundary of the square
$s_{i}^{2}$, then it is included into the boundaries of four edges 
$s_{m}^{1}$ two of which are included into the boundary of the square 
$s_{i}^{2}$. The condition (\ref{3.1}) is fulfilled again.

Let $G(M_{1}^{\prime },M_{2}^{\prime };M_{1},M_{2})$ be a rectangular
lattice on the plane. We identify the opposite sides of the entire 
rectangular. The obtained graph is denoted by 
$\tilde{G} (M_{1}^{\prime },M_{2}^{\prime};M_{1},M_{2})$. It is called
the rectangular lattice on the torus. Let a graph $G$ be embedded in
the graph $\tilde{G} (M_{1}^{\prime },M_{2}^{\prime};M_{1},M_{2})$. 
Let all the vertices from the boundaries of all the edges of a graph
$G$ be included into the set of the vertices of a graph $G$. The
cell complex $P(G)$ is the set consisting of vertices, edges and faces.
The sets of vertices and edges of $P(G)$ coincide with the sets of
vertices and edges of a graph $G$. The set of faces of $P(G)$ consists
of all the faces whose all edges of the boundaries are included into a
graph $G$. The cells $s_{i}^{p}$ and the incidence numbers 
$(s_{i}^{p}:s_{j}^{p - 1})$ are defined similarly to the plane case. For
any pair of the cells $s_{i}^{2}$, $s_{j}^{0}$ the incidence numbers
satisfy the condition (\ref{3.1}).

We suppose that a graph $G$ is embedded in a rectagular lattice on the
plane or on the torus. A cochain $c^{p}$ of the complex $P(G)$ with the 
coefficients in the group ${\bf Z}_{2}^{add}$ is a function on the $p$
dimensional cells taking values in the group ${\bf Z}_{2}^{add}$.
Usually the cell orientation  is considered and the cochains are the 
antisymmetric functions: $c^{p}(- s^{p}) = - c^{p}(s^{p})$. However,
$- 1 = 1 \, \hbox{mod} \, 2$ and we can neglect the cell orietation for
the coefficients in the group ${\bf Z}_{2}^{add}$. The cochains form an
Abelian group
\begin{equation}
\label{3.2}
(c^{p} + c^{\prime p})(s_{i}^{p}) = c^{p}(s_{i}^{p}) + c^{\prime p}(s_{i}^{p})
\, \hbox{mod} \, 2.
\end{equation} 
It is denoted by $C^{p}(P(G),{\bf Z}_{2}^{add})$. The mapping
\begin{equation}
\label{3.3}
\partial c^{p}(s_{i}^{p - 1}) = \sum_{j} 
(s_{j}^{p}:s_{i}^{p - 1}) c^{p}(s_{j}^{p}) \, \hbox{mod} \, 2
\end{equation}
defines the homomorphism of the group $C^{p}(P(G),{\bf Z}_{2}^{add})$
into the group $C^{p - 1}(P(G),{\bf Z}_{2}^{add})$. It is called the
boundary operator. The mapping
\begin{equation}
\label{3.4}
\partial^{\ast} c^{p}(s_{i}^{p + 1}) = \sum_{j} 
(s_{i}^{p + 1}:s_{j}^{p}) c^{p}(s_{j}^{p}) \, \hbox{mod} \, 2
\end{equation}
defines the homomorphism of the group $C^{p}(P(G),{\bf Z}_{2}^{add})$
into the group $C^{p + 1}(P(G),{\bf Z}_{2}^{add})$. It is called the
coboundary operator. The condition (\ref{3.1}) implies 
$\partial \partial = 0$, $\partial^{\ast} \partial^{\ast}  = 0$.
The kernel $Z_{p}(P(G),{\bf Z}_{2}^{add})$ of the homomorphism (\ref{3.3})
on the group $C^{p}(P(G),{\bf Z}_{2}^{add})$ is called the group of cycles of 
the complex $P(G)$ with the coefficients in the group ${\bf Z}_{2}^{add}$. 
The image $B_{p}(P(G),{\bf Z}_{2}^{add})$ of the homomorphism (\ref{3.3}) 
in the group $C^{p}(P(G),{\bf Z}_{2}^{add})$ is called the group of
boundaries of the complex $P(G)$ with the coefficients in the group
${\bf Z}_{2}^{add}$. Since $\partial \partial = 0$, the group
$B_{p}(P(G),{\bf Z}_{2}^{add})$ is the subgroup of the group 
$Z_{p}(P(G),{\bf Z}_{2}^{add})$. Analogously, for the coboundary operator 
$\partial^{\ast } $ the group of cocycles $Z^{p}(P(G),{\bf Z}_{2}^{add})$
and the group of coboundaries $B^{p}(P(G),{\bf Z}_{2}^{add})$ are
defined.

It is possible to introduce the bilinear form on 
$C^{p}(P(G),{\bf Z}_{2}^{add})$:
\begin{equation}
\label{3.5}
\langle f^{p},g^{p} \rangle = \sum_{i} f^{p}(s_{i}^{p})g^{p}(s_{i}^{p})
\, \hbox{mod} \, 2.
\end{equation}
The definitions (\ref{3.3}) and (\ref{3.4}) imply
\begin{eqnarray}
\label{3.6}
\langle f^{p},\partial^{\ast} g^{p - 1} \rangle =
\langle \partial f^{p},g^{p - 1} \rangle \nonumber \\
\langle f^{p},\partial g^{p + 1} \rangle =
\langle \partial^{\ast} f^{p},g^{p + 1} \rangle .
\end{eqnarray}

Let a cochain $\sigma \in C^{0}(P(G),{\bf Z}_{2}^{add})$. Let the
energy be expressed in the form
\begin{equation}
\label{3.7}
H^{\prime }(\partial^{\ast} \sigma ) = \sum_{s_{i}^{1} \in P(G)}
h_{i}(\partial^{\ast} \sigma (s_{i}^{1}))
\end{equation}
where $h_{i}(\epsilon )$ is an arbitrary function on the group 
${\bf Z}_{2}^{add}$:
\begin{equation}
\label{3.8}
h_{i}(\epsilon ) = D_{i} - E_{i}(- 1)^{\epsilon}
\end{equation}
and the constants
\begin{eqnarray}
\label{3.9}
D_{i} = 1/2 (h_{i}(1) + h_{i}(0)) \nonumber \\
E_{i} = 1/2 (h_{i}(1) - h_{i}(0)).
\end{eqnarray}
The substitution of the equality (\ref{3.8}) into the equality (\ref{3.7})
gives
\begin{equation}
\label{3.10}
H^{\prime }(\partial^{\ast} \sigma ) = \sum_{s_{i}^{1} \in P(G)} D_{i} 
+ H(\partial^{\ast} \sigma )
\end{equation} 
where the function
\begin{equation}
\label{3.11}
H(\partial^{\ast} \sigma ) = - \sum_{s_{i}^{1} \in P(G)} E_{i}
(- 1)^{\partial^{\ast} \sigma (s_{i}^{1})}
\end{equation}
is called the energy for the Ising model with zero magnetic field.
The number $E_{i} = E(s_{i}^{1})$ is the interaction energy attached 
to the edge $s_{i}^{1}$. The edge $s_{i}^{1}$ is given by its initial
vertex and by its direction. For example, the edges of a rectangular
lattice on the plane may be horizontal or vertical. If the interaction
energy $E_{i} = E(s_{i}^{1})$ is independent of the initial vertex of 
the edge $s_{i}^{1}$, the Ising model is called homogeneous. If the
interaction energy $E_{i} = E(s_{i}^{1})$ is independent of the direction 
of the edge $s_{i}^{1}$, the Ising model is called isotropic.

The equality (\ref{3.10}) implies
\begin{eqnarray}
\label{3.12}
Z_{G}^{\prime} = \sum_{\sigma \in C^{0}(P(G),Z_{2}^{add})}
\exp \{ - \beta H^{\prime }(\partial^{\ast} \sigma )\} = \nonumber \\
Z_{G}\exp \{ - \beta \sum_{s_{i}^{1} \in P(G)} D_{i}\}
\end{eqnarray}
where the function
\begin{equation}
\label{3.13}
Z_{G} = \sum_{\sigma \in C^{0}(P(G),Z_{2}^{add})}
\exp \{ - \beta H(\partial^{\ast} \sigma )\}
\end{equation}
is called the partition function of Ising model.

Let the cochain $\chi \in C^{0}(P(G),{\bf Z}_{2}^{add})$ take the value
$1$ at the vertices $x_{1}$,...,$x_{m}$ and be equal to $0$ at all
other vertices of the graph $G$. The correlation function at the vertices
$x_{1}$,...,$x_{m}$ of the lattice $G$ is the function
\begin{eqnarray}
\label{3.14}
W_{G}(\chi ) = (Z_{G}^{\prime })^{- 1}
\sum_{\sigma \in C^{0}(P(G),Z_{2}^{add})} 
(- 1)^{\langle \chi ,\sigma \rangle }
\exp \{ - \beta H^{\prime }(\partial^{\ast} \sigma )\} = \nonumber \\
(Z_{G})^{- 1}
\sum_{\sigma \in C^{0}(P(G),Z_{2}^{add})} 
(- 1)^{\langle \chi ,\sigma \rangle }
\exp \{ - \beta H(\partial^{\ast} \sigma )\} .
\end{eqnarray}
{\bf Proposition 3.1.} {\it The partition fuction of Ising model on the
graph} $G$
\begin{eqnarray}
\label{3.15}
Z_{G} = 2^{\# (VG)} 
\bigl( \prod_{s_{i}^{1} \in P(G)} \cosh \beta E(s_{i}^{1})\bigr)
\times \nonumber \\
\sum_{\xi^{1} \in Z_{1}(P(G),Z_{2}^{add})}
\prod_{s_{i}^{1} \in P(G)} 
(\tanh \beta E(s_{i}^{1}))^{1/2(1 - (- 1)^{\xi^{1} (s_{i}^{1})})}
\end{eqnarray}
{\it where $\# (VG)$ is the total number of the vertices of the graph} $G$.

{\it The correlation function of Ising model on the graph} $G$
\begin{eqnarray}
\label{3.16}
W_{G}(\chi ) = (Z_{G})^{- 1} 2^{\# (VG)} 
\bigl( \prod_{s_{i}^{1} \in P(G)} \cosh \beta E(s_{i}^{1})\bigr)
\times \nonumber \\
\sum_{{\xi^{1} \in C^{1}(P(G),Z_{2}^{add}),} \atop {\partial \xi^{1} = \chi }}
\prod_{s_{i}^{1} \in P(G)} 
(\tanh \beta E(s_{i}^{1}))^{1/2(1 - (- 1)^{\xi^{1} (s_{i}^{1})})}.
\end{eqnarray}
{\it Proof.} The definition (\ref{3.11}) implies
\begin{equation}
\label{3.17}
\exp \{ - \beta H(\sigma^{1} )\} = \prod_{s_{i}^{1} \in P(G)}
\exp \{ \beta E(s_{i}^{1})(- 1)^{\sigma^{1} (s_{i}^{1})}\}
\end{equation}
where $\sigma^{1} \in C^{1}(P(G),{\bf Z}_{2}^{add})$. It is easy to
verify that for $\epsilon = 0,1$
\begin{equation}
\label{3.18}
\exp \{ \beta E(s_{i}^{1})(- 1)^{\epsilon }\} =
(\cosh \beta E(s_{i}^{1})) \sum_{\xi = 0,1} (- 1)^{\xi \epsilon }
(\tanh \beta E(s_{i}^{1}))^{1/2(1 - (- 1)^{\xi })}.
\end{equation}
The relations (\ref{3.17}), (\ref{3.18}) imply
\begin{eqnarray}
\label{3.19}
\exp\{ - \beta H(\sigma^{1} )\} =
\bigl( \prod_{s_{i}^{1} \in P(g)} \cosh \beta E(s_{i}^{1})\bigr)
\times \nonumber \\
\sum_{\xi^{1} \in C^{1}(P(G),Z_{2}^{add})} 
(- 1)^{\langle \xi^{1},\sigma^{1} \rangle } \prod_{s_{i}^{1} \in P(g)}
(\tanh \beta E(s_{i}^{1}))^{1/2(1 - (- 1)^{\xi^{1} (s_{i}^{1})})}.
\end{eqnarray} 
The substitution of the equality (\ref{3.19}) into the definition
(\ref{3.13}), the first relation (\ref{3.6}) and the relation
\begin{equation}
\label{3.20}
\sum_{\xi = 0,1} (- 1)^{\xi \epsilon } =
\left\{ {2, \hskip 1cm \epsilon = 0,} \atop {0, \hskip 1cm \epsilon = 1,}
\right.
\end{equation}
give the equality (\ref{3.15}). The substitution of the equality
(\ref{3.19}) into the definition (\ref{3.14}), the first relation
(\ref{3.6}) and the relation (\ref{3.20}) give the equality (\ref{3.16}).
The proposition is proved.

Here we used the definitions and the methods of the paper \cite{17} .
The equality (\ref{3.15}) was proved for the first time in the paper 
\cite{16} . The equality (\ref{3.16}) for homogeneous isotropic Ising
model was proved in the paper \cite{19} .

The equality (\ref{3.16}) implies that the correlation function
$W_{G}(\chi )$ is not zero only for the cochains 
$\chi \in B_{0}(P(G),{\bf Z}_{2}^{add})$. Therefore the cochain $\chi $
takes the value $1$ only at the ends of the broken lines. Any broken
line has two ends. Hence the cochain $\chi $ takes the value $1$ at
even number of vertices.

\section{Free Boundary Conditions}
\setcounter{equation}{0}

\noindent We denote the one dimensional cell $s_{i}^{1}$ as the
non -- oriented edge $e_{i}$ corresponding with two oppositely
oriented edges ${\bf e}_{i}$ and ${\bf e}_{i}^{- 1}$. The interaction
energy of Ising model is denoted by $E(s_{i}^{1}) = E(e_{i}) =
E({\bf e}_{i}) = E({\bf e}_{i}^{- 1})$. The equality (\ref{3.15}) may
be rewritten in the form
\begin{equation}
\label{4.1}
Z_{G} = 2^{\# (VG)}\bigl( \prod_{e \in P(G)} \cosh \beta E(e)\bigr) 
Z_{r,G}
\end{equation}
where
\begin{equation}
\label{4.2}
Z_{r,G} = \sum_{\xi^{1} \in Z_{1}(P(G),Z_{2}^{add})} {\bf u}^{\xi^{1} },
\end{equation}
\begin{equation}
\label{4.3}
{\bf u}^{\xi^{1} } = \prod_{e \in P(G)} 
u(e)^{1/2(1 - (- 1)^{\xi^{1} (e)})} ,
\end{equation}
\begin{equation}
\label{4.4}
u(e) = u({\bf e}) = u({\bf e}^{- 1}) = \tanh \beta E(e).
\end{equation}

Let a graph $G$ be embedded in a rectangular lattice ${\bf Z}^{\times 2}$
on the plane. Let with any pair ${\bf e}_{1}$, ${\bf e}_{2}$ of the
oriented edges of a graph $G$ such that $f({\bf e}_{1}) = b({\bf e}_{2})$,
$b({\bf e}_{1}) \neq f({\bf e}_{2})$ there correspond the number
\begin{equation}
\label{4.5}
\rho ({\bf e}_{1};{\bf e}_{2}) = 
\exp \{ i/2\hat{({\bf e}_{1},{\bf e}_{2})} \}
\end{equation}
where $\hat{({\bf e}_{1},{\bf e}_{2})} $ is the radian measure of the
angle between the direction of the oriented edge ${\bf e}_{1}$ and 
the direction of the oriented edge ${\bf e}_{2}$. Due to the equalities
(\ref{2.17}) and (\ref{4.5}) with any reduced closed path $C$ on
the graph $G$ there corresponds the number $\rho (C) = \exp \{ i/2\phi (C)\}$ 
where $\phi (C)$ is the total angle through which the tangent vector
of the path $C$ turns along the path $C$.

For a graph $G$ embedded in the rectangular lattice ${\bf Z}^{\times 2}$
on the plane the estimate (\ref{2.30}) has the following form
\begin{equation}
\label{4.6}
|u({\bf e})| = |\tanh \beta E({\bf e})| < 1/3.
\end{equation}
{\bf Theorem 4.1.} {\it Let a finite graph} $G$ {\it be embedded in
the rectangular lattice} ${\bf Z}^{\times 2}$ {\it on the plane and
let the estimate} (\ref{4.6}) {\it be fulfilled. Then for the reduced
partition function} (\ref{4.2})
\begin{equation}
\label{4.7}
Z_{r,G} = \exp \{ - 1/2 \sum_{C \in RC(G)} |C|^{- 1}{\bf u}^{C}
\rho (C)\}
\end{equation}
{\it where} $C$ {\it runs over the set} $RC(G)$ {\it of all reduced
closed paths on a graph} $G$ {\it and the number} ${\bf u}^{C}$
{\it is defined by the relations} (\ref{2.18}), (\ref{4.4}).

\noindent {\it Proof.} Due to the paper \cite{12}
\begin{equation}
\label{4.8}
(Z_{r,G})^{2} = \det (I - T({\bf u},\rho ))
\end{equation}
where $(\# ({\bf E}G))\times (\# ({\bf E}G))$ -- matrix $T({\bf u},\rho )$
is defined by the equalities (\ref{2.32}), (\ref{4.4}) and (\ref{4.5}).
In view of the definition the angle $\phi (C) = 2\pi k$ where $k$ is
an integer. Hence the number $\rho (C) = \exp \{ i/2\phi (C)\} $ is
real. Now the equality (\ref{4.7}) follows from the equalities 
(\ref{2.36}) and (\ref{4.8}). The theorem is proved.

For the homogeneous two dimensional Ising model the interaction 
energy $E({\bf e})$ does not depend on an initial vertex of an edge
${\bf e}$. Let us denote $E_{1}$ ($E_{2}$) the interaction energy
$E({\bf e})$ for horizotally (vertically) directed edges ${\bf e}$.
The relations (\ref{4.1}) -- (\ref{4.4}) imply for the homogeneous
Ising model
\begin{equation}
\label{4.9}
Z_{G(M_{1}^{\prime },M_{2}^{\prime };M_{1},M_{2})} =
Z_{G(1,1;M_{1} - M_{1}^{\prime } + 1,M_{2} - M_{2}^{\prime } + 1)}.
\end{equation}
{\bf Theorem 4.2.} {\it Let for the interaction energy of the homogeneous
two dimensional Ising model the estimate} (\ref{4.6}) {\it be valid.
Then for the partition function} (\ref{4.1}) {\it of the homogeneous
Ising model on the rectangular lattice on the plane}
\begin{eqnarray}
\label{4.10}
\lim_{M_{i} \rightarrow \infty \, i = 1,2} 
(M_{1}M_{2})^{- 1} \ln Z_{G(1,1;M_{1},M_{2})} = \nonumber \\
\ln (2\cosh \beta E_{1}\cosh \beta E_{2}) + 
1/2 (2\pi )^{- 2} \int_{0}^{2\pi } d\theta_{1} \int_{0}^{2\pi } d\theta_{2}
\nonumber \\
\ln [(1 + z_{1}^{2})(1 + z_{2}^{2}) - 2z_{1}(1 - z_{2}^{2})\cos \theta_{1} -
2z_{2}(1 - z_{1}^{2})\cos \theta_{2} ]
\end{eqnarray}
{\it where the variables} $z_{i} = \tanh \beta E_{i}$, $i = 1,2$.

\noindent {\it Proof.} It follows from the equalities (\ref{4.1}) and
(\ref{4.7}) for the homogeneous Ising model on the rectangular lattice
that 
\begin{eqnarray}
\label{4.11}
\lim_{M_{i} \rightarrow \infty \, i = 1,2} 
(M_{1}M_{2})^{- 1} \ln Z_{G(1,1;M_{1},M_{2})} = \nonumber \\
\ln (2\cosh \beta E_{1}\cosh \beta E_{2}) - 
\lim_{M_{i} \rightarrow \infty \, i = 1,2} 
(2M_{1}M_{2})^{- 1} \sum_{C \in RC(G)} |C|^{- 1}{\bf u}^{C} \rho (C)
\end{eqnarray}
where $C$ runs over the set $RC(G)$ of all reduced closed paths on the
graph $G(1,1;M_{1},M_{2})$.

The total number of all reduced closed paths of the length $l$ 
with the initial vertex $(0,0)$ on the lattice ${\bf Z}^{\times 2}$ is less
than $4\cdot 3^{l - 1}$. Due to definitions (\ref{2.17}), (\ref{4.5})
the number $|\rho (C)| = 1$. Hence the estimate (\ref{4.6}) implies
the absolute convergence of the series in the right hand side of the
equality (\ref{4.11}) for finite $M_{1}$,$M_{2}$.

The interaction energy $E({\bf e})$ does not depend on the initial
vertex of the oriented edge ${\bf e}$. Therefore due to definitions
(\ref{2.17}), (\ref{4.5}), (\ref{2.18}) and (\ref{4.4}) the number
$|C|^{- 1}{\bf u}^{C}\rho (C)$ does not depend on the initial vertex
of the path $C$. Hence 
\begin{equation}
\label{4.12}
\sum_{C \in RC(G)} |C|^{- 1}{\bf u}^{C} \rho (C) =
\sum_{C_{0} \in RC_{0}(Z^{\times 2})} N(C_{0})|C_{0}|^{- 1}{\bf u}^{C_{0}} 
\rho (C_{0})
\end{equation}
where $C_{0}$ runs over the set $RC_{0}({\bf Z}^{\times 2})$ of all
reduced closed paths  with the starting point at the vertex $(0,0)$ 
on the lattice ${\bf Z}^{\times 2}$. The number $N(C_{0})$ is the
total number of shifted paths $C_{0}$ on the graph $G(1,1;M_{1},M_{2})$.
If $|C_{0}| \leq M_{1}$, $|C_{0}| \leq M_{2}$, the following estimate
is valid
\begin{equation}
\label{4.13}
(M_{1} - |C_{0}|)(M_{2} - |C_{0}|) \leq N(C_{0}) \leq M_{1}M_{2}.
\end{equation}
The estimate (\ref{4.6}) implies the absolute convergence of the series
\begin{equation}
\label{4.14}
\sum_{C_{0} \in RC_{0}(Z^{\times 2})} |C_{0}|^{k}{\bf u}^{C_{0}} \rho (C_{0})
\end{equation}
for $k = 0,\pm 1$. Thus it follows from the equalities (\ref{4.11}),
(\ref{4.12}) and estimates (\ref{4.13}) that
\begin{eqnarray}
\label{4.15}
\lim_{M_{i} \rightarrow \infty \, i = 1,2} 
(M_{1}M_{2})^{- 1} \ln Z_{G(1,1;M_{1},M_{2})} = \nonumber \\
\ln (2\cosh \beta E_{1}\cosh \beta E_{2}) -  
1/2\sum_{C_{0} \in RC_{0}(Z^{\times 2})} |C_{0}|^{- 1}{\bf u}^{C_{0}} 
\rho (C_{0}).
\end{eqnarray}

Let us rewrite the expression (\ref{4.15}) in more traditional form.
On the rectangular lattice $\tilde{G} (0,0;M_{1},M_{2})$ on the torus 
the number $u({\bf e})$ is defined by the equality (\ref{4.4}) and the
number $\rho ({\bf e}_{1};{\bf e}_{2})$ is defined by the equality
(\ref{4.5}) taking into account the identification of the vertices
and edges in the graph $\tilde{G} (0,0;M_{1},M_{2})$. The
$(4M_{1}M_{2}) \times (4M_{1}M_{2})$ -- matrix $T({\bf u},\rho )$ is
defined by the relation (\ref{2.32}). The equality (\ref{2.36}) implies
\begin{equation}
\label{4.16}
(2M_{1}M_{2})^{- 1} \ln [\det (I - T({\bf u},\rho ))] =
- (2M_{1}M_{2})^{- 1} \sum_{C \in RC(\tilde{G} )}
|C|^{- 1}{\bf u}^{C}\rho (C)
\end{equation}
where $C$ runs over the set $RC(\tilde{G} )$ of all reduced closed 
paths on the graph $\tilde{G} (0,0;M_{1},M_{2})$.

For the homogeneous Ising model the number $|C|^{- 1}{\bf u}^{C}\rho (C)$ 
does not depend on the starting point of the path $C$. The graph 
$\tilde{G} (0,0;M_{1},M_{2})$ is invariant under the shifts. Any reduced 
closed path $C$ is a shifted reduced closed path $C_{0}$ where a path 
$C_{0}$ has the starting point at the vertex $(0,0)$. Therefore the 
equality (\ref{4.16}) implies
\begin{equation}
\label{4.17}
(2M_{1}M_{2})^{- 1} \ln [\det (I - T({\bf u},\rho ))] =
- 1/2 \sum_{C_{0} \in RC_{0}(\tilde{G} )}
|C_{0}|^{- 1}{\bf u}^{C_{0}}\rho (C_{0})
\end{equation}
where $C_{0}$ runs over the set $RC_{0}(\tilde{G} )$ of all reduced closed 
paths  with the starting point at the vertex $(0,0)$ on the graph 
$\tilde{G} (0,0;M_{1}M_{2})$. When $M_{i} \rightarrow \infty $, $i = 1,2$, 
the terms in the series (\ref{4.17}) corresponding with the long paths 
connecting the vertices $(0,k)$ and $(M_{1},k)$ or the $(k,0)$ and 
$(k,M_{2})$, etc tend to zero due to the estimate (\ref{4.6}). Hence
\begin{equation}
\label{4.18}
\lim_{M_{i} \rightarrow \infty \, i = 1,2}
(2M_{1}M_{2})^{- 1} \ln [\det (I - T({\bf u},\rho ))] =
- 1/2 \sum_{C_{0} \in RC_{0}(Z^{\times 2})}
|C_{0}|^{- 1}{\bf u}^{C_{0}}\rho (C_{0})
\end{equation}
where $C_{0}$ runs over the set $RC_{0}({\bf Z}^{\times 2})$ of all reduced 
closed paths  with the starting point at the vertex $(0,0)$ on the 
lattice ${\bf Z}^{\times 2}$.

Let us calculate the determinant of the matrix $I - T({\bf u},\rho )$.
The vertices of the graph $\tilde{G} (0,0;M_{1},M_{2})$ are defined
by the vectors ${\bf j} \in {\bf Z}^{\times 2}$, $1 \leq j_{1} \leq M_{1}$,
$1 \leq j_{2} \leq M_{2}$. The oriented edge ${\bf e}$ of the graph
$\tilde{G} (0,0;M_{1},M_{2})$ is defined by its initial vertex
$b({\bf e}) = {\bf j}$ and by its direction: the unit vector ${\bf v}$.
The unit vector ${\bf v}$ is one of four vectors: $(\pm 1,0)$, $(0,\pm 1)$.
Thus an oriented edge ${\bf e}$ is a pair $({\bf j},{\bf v})$. Due to
definitions (\ref{2.32}), (\ref{4.4}) and (\ref{4.5})
\begin{eqnarray}
\label{4.19}
T({\bf u},\rho )_{(j,v),(j^{\prime },v^{\prime })} = u(({\bf j},{\bf v}))
\delta_{j_{1} + v_{1} - j_{1}^{\prime },\, M_{1}Z}
\delta_{j_{2} + v_{2} - j_{2}^{\prime },\, M_{2}Z} \times \nonumber \\
(1 - 
\delta_{v_{1} + v_{1}^{\prime },\, 0} \delta_{v_{2} + v_{2}^{\prime },\, 0} )
\exp \{ i/2\hat{({\bf v},{\bf v}^{\prime })} \}
\end{eqnarray}
where Kronecker symbol
\begin{equation}
\label{4.20}
\delta_{j,\, M_{k}Z} = M_{k}^{- 1} \sum_{l = 1}^{M_{k}}
\exp \{ i2\pi M_{k}^{- 1}jl\}.
\end{equation}
The symbol (\ref{4.20}) equals $1$ if $j = M_{k}l$ where $l$ is an integer
and it equals $0$ if $j \neq M_{k}l$.

For the homogeneous Ising model the number $u(({\bf j},{\bf v}))$
does not depend on the vector ${\bf j}$. We denote it by u({\bf v}).
The equalities (\ref{4.19}) and (\ref{4.20}) imply
\begin{equation}
\label{4.21}
(I - T({\bf u},\rho ))_{(j,v), \, (j^{\prime },v^{\prime })} =
(CBC^{- 1})_{(j,v), \, (j^{\prime },v^{\prime })}
\end{equation}
where the matrices
\begin{equation}
\label{4.22}
C_{(j,v),\, (j^{\prime },v^{\prime })} = \delta_{v_{1},\, v_{1}^{\prime }}
\delta_{v_{2},\, v_{2}^{\prime }} (M_{1}M_{2})^{- 1/2}
\exp \{ i2\pi (M_{1}^{- 1}j_{1}j_{1}^{\prime } + 
M_{2}^{- 1}j_{2}j_{2}^{\prime })\} ,
\end{equation}
\begin{eqnarray}
\label{4.23}
B_{(j,v),\, (j^{\prime },v^{\prime })} =
\delta_{j_{1},\, j_{1}^{\prime }} \delta_{j_{2},\, j_{2}^{\prime }}
B({\bf j})_{v,\, v^{\prime }} , \nonumber \\ 
B({\bf j})_{v,\, v^{\prime }} = 
\delta_{v_{1},\, v_{1}^{\prime }} \delta_{v_{2},\, v_{2}^{\prime }} -
\nonumber \\
\exp \{ i2\pi (M_{1}^{- 1}v_{1}j_{1} + M_{2}^{- 1}v_{2}j_{2}) \} 
u({\bf v})\exp \{ i/2\hat{({\bf v},{\bf v}^{\prime })} \}
(1 - 
\delta_{v_{1} ,\, - v_{1}^{\prime }} \delta_{v_{2} ,\, - v_{2}^{\prime }} ).
\end{eqnarray}
The matrix $B_{(j,v),\, (j^{\prime },v^{\prime })}$ is diagonal for
the vectors ${\bf j}$,${\bf j}^{\prime }$. The second relation 
(\ref{4.23}) defines $4\times 4$ -- matrix $B({\bf j})_{v,\, v^{\prime}}$
for any vector ${\bf j} \in {\bf Z}^{\times 2}$, $1 \leq j_{1} \leq M_{1}$,
$1 \leq j_{2} \leq M_{2}$. It follows from the relations (\ref{4.21})
and (\ref{4.23}) that
\begin{equation}
\label{4.24}
\det (I - T({\bf u},\rho )) = \prod_{j_{1} = 1}^{M_{1}}
\prod_{j_{2} = 1}^{M_{2}} \det B({\bf j}).
\end{equation}
Due to relations (\ref{4.4}) $u((\pm 1,0)) = \tanh \beta E_{1} = z_{1}$,
$u(0,\pm 1) = \tanh \beta E_{2} = z_{2}$. By using the definition
(\ref{4.23}) it is possible to calculate
\begin{equation}
\label{4.25}
\det B({\bf j}) = (1 + z_{1}^{2})(1 + z_{2}^{2}) -
2z_{1}(1 - z_{2}^{2})\cos 2\pi M_{1}^{- 1}j_{1} -
2z_{2}(1 - z_{1}^{2})\cos 2\pi M_{2}^{- 1}j_{2}.
\end{equation}
The substitution of the equality (\ref{4.25}) into the relation 
(\ref{4.24}) yields
\begin{eqnarray}
\label{4.26}
\det (I - T({\bf u},\rho )) = \nonumber \\
\prod_{j_{1} = 1}^{M_{1}} \prod_{j_{2} = 1}^{M_{2}}
[(1 + z_{1}^{2})(1 + z_{2}^{2}) -
2z_{1}(1 - z_{2}^{2})\cos 2\pi M_{1}^{- 1}j_{1} -
2z_{2}(1 - z_{1}^{2})\cos 2\pi M_{2}^{- 1}j_{2}].
\end{eqnarray}
The equalities (\ref{4.15}), (\ref{4.18}) and (\ref{4.26}) imply the
equality (\ref{4.10}). The theorem is proved.

Let a cochain $\xi^{1} \in C^{1}(P(G),{\bf Z}_{2}^{add})$. The support
$||\xi^{1} ||$ is the set of all non -- oriented edges of the graph
$G$ on which a cochain $\xi^{1} $ takes the value $1$. Let a cochain
$\chi \in C_{0}(P(G),{\bf Z}_{2}^{add})$. The support $||\xi^{1} ||$
is called $\chi $ -- connected if any connected component of the
support $||\xi^{1} ||$ contains the non -- oriented edges incident to
the vertices on which a cochain $\chi $ equals $1$. Let $i(||\xi^{1} ||)$
be the set of all non -- oriented edges incident to the vertices incident
to the edges of the support $||\xi^{1} ||$. By using the relations
(\ref{4.1}) -- (\ref{4.4}) we rewrite the correlation function (\ref{3.16})
in the following form
\begin{equation}
\label{4.27}
W_{G}(\chi ) = (Z_{r,G})^{- 1}
\sum_{{\xi^{1} \in C^{1}(P(G),Z_{2}^{add}), \, \partial \xi^{1} = \chi ,}
\atop {\chi - connected \, ||\xi^{1} ||}}
{\bf u}^{\xi^{1} } Z_{r,G \setminus i(||\xi^{1} ||)}
\end{equation}
where the graph $G \setminus i(||\xi^{1} ||)$ is obtained by deleting all
edges of the set $i(||\xi^{1} ||)$ from the graph $G$.

If $C$ is a closed path on the graph $G$, the support $||C||$ is the set
of all non -- oriented edges corresponding with the oriented edges from a
closed path $C$.

\noindent {\bf Theorem 4.3.} {\it Let the estimate} (\ref{4.6}) {\it be
valid. Let the interaction energy} $E({\bf e})$ {\it be non -- negative.
Let a cochain} $\chi \in C^{0}(P(G),{\bf Z}_{2}^{add})$ {\it be equal to} 
$1$ {\it on the finite number of the vertices. Then for the correlation 
function} (\ref{3.14}) {\it of the two dimensional Ising model with free 
boundary conditions} 
\begin{eqnarray}
\label{4.28}
\lim_{{M_{k} \rightarrow \infty ,\, M_{k}^{\prime } \rightarrow - \infty ,}
\atop {k = 1,2}} W_{G(M_{1}^{\prime },M_{2}^{\prime };M_{1},M_{2})} (\chi ) =
\nonumber \\
\sum_{{\xi^{1} \in C^{1}(P(G),Z_{2}^{add}),\, \partial \xi^{1} = \chi ,}
\atop {\chi - connected \, ||\xi^{1} ||}} {\bf u}^{\xi^{1} }
\exp \{ 1/2\sum_{{C \in RC(Z^{\times 2}),}
\atop {||C||\cap i(||\xi^{1} ||) \neq \emptyset }}
|C|^{- 1} {\bf u}^{C}\rho (C) \}
\end{eqnarray}
{\it where the number} ${\bf u}^{\xi^{1} }$ {\it is defined by the relations} 
(\ref{4.3}), (\ref{4.4}), {\it the number} ${\bf u}^{C}$ {\it is defined by
the relations} (\ref{2.18}), (\ref{4.4}) {\it and the number} $\rho (C)$
{\it is defined by the relations} (\ref{2.17}), (\ref{4.5}).

\noindent {\it Proof.} Let us consider the equality (\ref{4.27}) for
the graph $G = G(M_{1}^{\prime },M_{2}^{\prime };M_{1},M_{2})$. 
Theorem 4.1 impiles
\begin{equation}
\label{4.29}
(Z_{r,G})^{- 1}Z_{r,G \setminus i(||\xi^{1} ||)} =
\exp \{ 1/2\sum_{{C \in RC(G),}
\atop {||C||\cap i(||\xi^{1} ||) \neq \emptyset }}
|C|^{- 1} {\bf u}^{C}\rho (C) \} .
\end{equation}
The total number of all reduced closed paths of the length $l$ with
fixed initial vertex on the lattice ${\bf Z}^{\times 2}$ is less than
$4\cdot 3^{l - 1}$. Due to the estimate (\ref{4.6}) the series (\ref{4.29})
is absolutely convergent for $G \rightarrow {\bf Z}^{\times 2}$. Thus
every term of the sum (\ref{4.27}) for the graph
$G = G(M_{1}^{\prime },M_{2}^{\prime};M_{1},M_{2})$ converges to the
term of the series (\ref{4.28}) when $G \rightarrow {\bf Z}^{\times 2}$.
The equality (\ref{4.28}) will be proved if the absolute convergence
of the series (\ref{4.28}) is proved when the estimate (\ref{4.6}) is
valid.

The correlation function (\ref{4.27}) is not zero only for the cochains
$\chi \in B_{0}(P(G),{\bf Z}_{2}^{add})$ taking the value $1$ at the
even number of vertices ${\bf m}_{1},...,{\bf m}_{2k}$ of the graph
$G \subset {\bf Z}^{\times 2}$. Let a cochain 
$\xi^{1} \in C^{1}(P(G),{\bf Z}_{2}^{add})$ satisfy the condition
$\partial \xi^{1} = \chi $. Then the vertex ${\bf m}_{1}$ is incident
to one or three non -- oriented edges on which a cochain $\xi^{1} $ 
takes the value $1$. We take one such edge. It corresponds with the 
oriented edge $({\bf m}_{1},{\bf v}_{1})$. If 
${\bf m}_{1} + {\bf v}_{1} = {\bf m}_{j_{1}}$, then we have constructed
the path connecting the vertices ${\bf m}_{1}$ and ${\bf m}_{j_{1}}$.
If ${\bf m}_{1} + {\bf v}_{1}$ does not coincide with any vertices
${\bf m}_{2},...,{\bf m}_{2k}$, then due to the condition 
$\partial \xi^{1} = \chi $ the vertex ${\bf m}_{1} + {\bf v}_{1}$ is
incident to two or four non -- oriented edges on which a cochain $\xi^{1} $
takes the value $1$. One of these non -- oriented edges corresponds to the
oriented edge $({\bf m}_{1},{\bf v}_{1})$. We take another edge. It
corresponds with the oriented edge $({\bf m}_{1} + {\bf v}_{1},{\bf v}_{2})$,
${\bf v}_{2} \neq - {\bf v}_{1}$. By repeating this process we obtain
the path $P_{1} = (({\bf m}_{1}^{\prime },{\bf v}_{1}),...,
({\bf m}_{q_{1}}^{\prime },{\bf v}_{q_{1}}))$ 
where ${\bf m}_{1}^{\prime } = {\bf m}_{1}$, 
${\bf m}_{i + 1}^{\prime } = {\bf m}_{i}^{\prime } + {\bf v}_{i}$,
$i = 1,...,q_{1} - 1$, 
${\bf m}_{q_{1}}^{\prime } + {\bf v}_{q_{1}} = {\bf m}_{j_{1}}$ and
${\bf m}_{j_{1}}$ is one of the vertices ${\bf m}_{2},...,{\bf m}_{2k}$
on which the cochain $\chi $ takes the value $1$. Any non -- oriented
edge may correspond with only one oriented edge from the path $P_{1}$.
Let the cochain $\xi^{1} [P_{1}] \in C^{1}(P(G),{\bf Z}_{2}^{add})$ 
equal $1$ on all non -- oriented edges corresponding with the oriented 
edges from the path $P_{1}$. It equals $0$ on all other non -- oriented edges 
from the graph $G$. By construction $\xi^{1} = \xi^{1} [P_{1}] + \eta^{1} $ 
where the supports of the cochains 
$\xi^{1} [P_{1}],\eta^{1} \in C^{1}(P(G),{\bf Z}_{2}^{add})$ 
do not intersect each other. By repeating this process we 
construct the paths $P_{1},...,P_{k}$ connecting the vertices 
${\bf m}_{i_{1}},...,{\bf m}_{i_{k}}$, $1 = i_{1} < i_{2} < \cdots < i_{k}$
with the vertices ${\bf m}_{j_{1}},...,{\bf m}_{j_{k}}$, $i_{l} < j_{l}$,
$l = 1,...,k$. Any non -- oriented edge may correspond with only one
oriented edge from the paths $P_{l}$, $l = 1,...,k$. These paths correspond 
with the cochains $\xi^{1} [P_{l}]$, $l = 1,...,k$, such that 
$\xi^{1} = \xi^{1} [P_{1}] + \cdots + \xi^{1} [P_{k}] + \eta^{1} $ where the 
supports of the cochains $\xi^{1} [P_{1}],...,\xi^{1} [P_{k}],\eta^{1} 
\in C^{1}(P(G),{\bf Z}_{2}^{add})$ do not intersect each other and 
$\eta^{1} \in Z_{1}(P(G),{\bf Z}_{2}^{add})$. This decomposition is not unique 
in general. Therefore not an equality but the estimate is valid
\begin{equation}
\label{4.30}
W_{G}(\chi ) \leq (Z_{r,G})^{- 1} 
\sum_{{\{ i_{l},j_{l}\},} \atop {l = 1,...,k}}
\sum_{{P_{l},} \atop {l = 1,...,k}} 
\bigl( \prod_{l = 1}^{k} {\bf u}^{P_{l}}\bigr)
Z_{r,G \setminus (\prod_{l = 1}^{k} ||P_{l}||)}
\end{equation}
where $\{ i_{l},j_{l}\} $ runs over the set of the subdivisions of
the numbers $1,...,2k$ into $k$ pairs: $1 = i_{1} < \cdots < i_{k}$,
$i_{l} < j_{l}$ , $l = 1,...,k$, the paths $P_{l}$, $l = 1,...,k$,
run over the set of the paths connecting the vertices ${\bf m}_{i_{l}}$
and ${\bf m}_{j_{l}}$, $l = 1,...,k$, any non -- oriented edge may 
correspond with only one oriented edge from the paths $P_{l}$, $l = 1,...,k$ 
and the graph $G \setminus (\prod_{l = 1}^{k} ||P_{l}||)$ is obtained 
from the graph $G$ by deleting all edges from the supports $||P_{l}||$, 
$l = 1,...,k$. Due to the definition (\ref{4.4}) the variable $u({\bf e})$ is 
non -- negative when the interaction energy $E({\bf e})$ is non -- negative.
For the non -- negative variables $u({\bf e})$ the definition (\ref{4.2}) 
implies the estimate
\begin{equation}
\label{4.31}
(Z_{r,G})^{- 1}Z_{r,G \setminus (\prod_{l = 1}^{k} ||P_{l}||)}
\leq 1.
\end{equation}
It follows from the estimates (\ref{4.30}), (\ref{4.31}) that
\begin{equation}
\label{4.32}
W_{G}(\chi ) \leq \sum_{{\{ i_{l},j_{l}\} ,} \atop {l = 1,...,k}}
\sum_{{P_{l},} \atop {l = 1,...,k}}
\prod_{l = 1}^{k} {\bf u}^{P_{l}}.
\end{equation}
The total number of the reduced closed paths of the length $l$ starting
at the fixed vertex on the lattice ${\bf Z}^{\times 2}$ is less than
$4 \cdot 3^{l - 1}$. Hence the estimate (\ref{4.6}) implies that for
$G \rightarrow {\bf Z}^{\times 2}$ the sum (\ref{4.32}) converges to 
the absolutely convergent series. This series majorizes the series 
(\ref{4.28}). Therefore the series (\ref{4.28}) is absolutely convergent 
when the estimate (\ref{4.6}) is valid. The theorem is proved.

Let a cycle $\xi^{1} \in Z_{1}(P(G),{\bf Z}_{2}^{add})$. By using the
arguments of Theorem 4.3 we can construct the closed paths $C_{1}$,...,
$C_{m}$ where the supports $||C_{1}||$,...,$||C_{m}||$ do not intersect
each other and any non -- oriented edge may correspond with only one 
oriented edge from the closed paths $C_{1}$,...,$C_{m}$. Any closed path 
on a rectangular lattice on the plane has an even number of the vertically
directed edges and it has an even number of the horizontally
directed edges. Let the interaction energy $E({\bf e})$ be
non -- negative for the vertically directed edges ${\bf e}$
and let it be non -- positive for the horizontally directed 
edges ${\bf e}$. Due to the definitions (\ref{4.3}), (\ref{4.4})
${\bf u}^{\xi^{1} } \geq 0$. Hence the inequality (\ref{4.31}) is
fulfilled in this case. Therefore it is possible to prove the absolute
convergence of the series (\ref{4.28}). Thus Theorem 4.3 is valid when
the interaction energy is non -- negative for the vertically directed 
edges and it is non -- positive for the horizontally directed
edges. Theorem 4.3 is valid also for the case when the interaction
energy is non -- positive for the vertically directed edges and
it is non -- negative for the horizontally directed edges. If
the interaction energy is non -- positive for all oriented edges, then
Theorem 4.3 is also valid. 

\section{Periodic Boundary Conditions}
\setcounter{equation}{0}

\noindent Let us consider the rectagular lattice 
$\tilde{G} (M_{1}^{\prime },M_{2}^{\prime };M_{1},M_{2})$ on the torus
introduced in Section 3. The number $\rho ({\bf e}_{1};{\bf e}_{2})$
is given by the relation (\ref{4.5}) taking into account 
the identification of the vertices and the edges in the graph 
$\tilde{G} (M_{1}^{\prime },M_{2}^{\prime };M_{1},M_{2})$. With every 
reduced closed path $C$ on the graph
$\tilde{G} (M_{1}^{\prime },M_{2}^{\prime };M_{1},M_{2})$ there
corresponds the number $\rho (C) = \exp \{ i/2\phi (C)\} $ defined by
the equality (\ref{2.17}). Here $\phi (C)$ is the total angle through
which the tangent vector of the path $C$ turns along the path $C$.

If the reduced closed path $C$ lies on the rectangular lattice 
${\bf Z}^{\times 2}$ and the number $\rho (C)$ is defined by the 
relations (\ref{2.17}), (\ref{4.5}), then due to \cite{20} the number
$\rho (C) = - (- 1)^{n(C)}$ where $n(C)$ is the total number of the
transversal self -- intersections of the path $C$. The papers \cite{5},
\cite{6}, \cite{12}, \cite{13} used this Whitney formula. Let us consider
the line $C$ connecting the vertices $(M_{1}^{\prime },k)$ and
$(M_{1},k)$ on the graph 
$\tilde{G} (M_{1}^{\prime },M_{2}^{\prime };M_{1},M_{2})$. The line has
no self -- intersections. It is easy to see that $\rho (C) = 1$. Thus 
Whitney formula is wrong for a torus in general. Therefore we can not
use the results of the papers \cite{5}, \cite{6}, \cite{12}, \cite{13}
for a torus.

Let us study the properties of the $L$ -- function (\ref{2.31}) for a
graph $G$ lying on the graph
$\tilde{G} (M_{1}^{\prime },M_{2}^{\prime };M_{1},M_{2})$. Let for the
reduced closed path 
$C = ({\bf e}_{1},...,{\bf e}_{p},{\bf e}_{p + 1},...,{\bf e}_{p + q})$
the vertices $b({\bf e}_{p + 1}) = b({\bf e}_{1})$. Then 
$C = C_{1}\cdot C_{2}$ where the closed paths 
$C_{1} = ({\bf e}_{1},...,{\bf e}_{p})$ and
$C_{2} = ({\bf e}_{p + 1},...,{\bf e}_{p + q})$ may be not reduced. Indeed,
if ${\bf e}_{1} = {\bf e}_{p}^{- 1}$, then the closed path $C_{1}$ is not
reduced. If ${\bf e}_{p + 1} = {\bf e}_{p + q}^{- 1}$, then the closed path
$C_{2}$ is not reduced. By the definition a reduced closed path does not 
contain the oppositely oriented edges ${\bf e}$, ${\bf e}^{- 1}$ if they
are subsequent or if they are the first and the last edges of the path.
A closed path is called completely reduced if it does not contained the 
oppositely oriented edges ${\bf e}$, ${\bf e}^{- 1}$ at any places. The 
multipliers $C_{1}$ and $C_{2}$ of any completely reduced closed path
$C = C_{1} \cdot C_{2}$ are also completely reduced closed paths. The set
of all completely reduced closed paths on the graph $G$ is denoted by
$CRC(G)$.

\noindent {\bf Theorem 5.1.} {\it Let a graph} $G$ {\it be embedded in the
rectangular lattice}
$\tilde{G} (M_{1}^{\prime },M_{2}^{\prime };M_{1},M_{2})$ {\it on the torus.
Let with any reduced closed path} $C$ {\it on the graph} $G$ {\it there
correspond the number} $\rho (C)$ {\it given by the relations} (\ref{2.17}),
(\ref{4.5}) {\it and the number} ${\bf u}^{C}$ {\it given by the relations}
(\ref{2.18}), (\ref{4.4}). {\it If the estimate} (\ref{4.6}) {\it is valid,
then}
\begin{equation}
\label{5.1}
\sum_{C \in RC(G)} |C|^{- 1}{\bf u}^{C}\rho (C) =
\sum_{C \in CRC(G)} |C|^{- 1}{\bf u}^{C}\rho (C).
\end{equation}
{\it In the left hand side of the equality} (\ref{5.1}) {\it the sum extends
over the set} $RC(G)$ {\it of all the reduced closed paths on the graph} $G$
{\it and in the right hand side of the equality} (\ref{5.1}) {\it the sum
extends over the set} $CRC(G)$ {\it of all the completely reduced closed paths
on the graph} $G$.

\noindent {\it Proof.} Let the reduced closed path $C = ({\bf e}_{1},...,
{\bf e}_{p},{\bf e},{\bf e}_{p + 1},...,{\bf e}_{p + q},{\bf e}^{- 1},
{\bf e}_{p + q + 1},...,{\bf e}_{p + q + r})$ contain the oppositely 
oriented edges ${\bf e}$ and ${\bf e}^{- 1}$. Then the closed path

\noindent
$C^{\prime } = ({\bf e}_{1},...,{\bf e}_{p},{\bf e},{\bf e}_{p + q}^{- 1},
...,{\bf e}_{p + 1}^{- 1},{\bf e}^{- 1},{\bf e}_{p + q + 1},...,
{\bf e}_{p + q + r})$ is also reduced.

The path length definition and the definitions (\ref{2.18}), (\ref{4.4})
imply
\begin{equation}
\label{5.2}
|C| = |C^{\prime }|, \, \, \, {\bf u}^{C} = {\bf u}^{C^{\prime }}.
\end{equation}
By using the definitions (\ref{2.17}), (\ref{4.5}) we get
\begin{equation}
\label{5.3}
\phi (C) = \phi_{1} + \phi_{2} + \phi_{3},
\end{equation}
\begin{equation}
\label{5.4}
\phi_{1} = \sum_{i = 1}^{p - 1} \hat{({\bf e}_{i},{\bf e}_{i + 1})} +
\hat{({\bf e}_{p},{\bf e})} ,
\end{equation}
\begin{equation}
\label{5.5}
\phi_{2} = \hat{({\bf e},{\bf e}_{p + 1})} + 
\sum_{i = p + 1}^{p + q - 1} \hat{({\bf e}_{i},{\bf e}_{i + 1})} +
\hat{({\bf e}_{p + q},{\bf e}^{- 1})} ,
\end{equation}
\begin{equation}
\label{5.6}
\phi_{3} = \hat{({\bf e}^{- 1},{\bf e}_{p + q + 1})} + 
\sum_{i = p + q + 1}^{p + q + r - 1} \hat{({\bf e}_{i},{\bf e}_{i + 1})} +
\hat{({\bf e}_{p + q + r},{\bf e}_{1})} .
\end{equation}
It is easy to verify the relation
\begin{equation}
\label{5.7}
\hat{({\bf e}_{1},{\bf e}_{2})} = - 
\hat{({\bf e}_{2}^{- 1},{\bf e}_{1}^{- 1})}
\end{equation}
for the oriented edges ${\bf e}_{1}$, ${\bf e}_{2}$ such that 
$f({\bf e}_{1}) = b({\bf e}_{2})$, $b({\bf e}_{1}) \neq f({\bf e}_{2})$.
The definitions (\ref{2.17}), (\ref{4.5}) and the relations (\ref{5.7})
imply
\begin{equation}
\label{5.8}
\phi (C^{\prime }) = \phi_{1} - \phi_{2} + \phi_{3}.
\end{equation}
Since the directions of the oriented edges ${\bf e}$ and ${\bf e}^{- 1}$
are opposite, due to the definition (\ref{5.5})
$\phi_{2} = (2k + 1)\pi $ where $k$ is an integer. Hence
$\exp \{ - i/2\phi_{2} \} = - \exp \{ i/2\phi_{2} \} $ and in view of the
relations (\ref{5.3}), (\ref{5.8})
\begin{equation}
\label{5.9}
\exp \{ i/2\phi (C)\} = - \exp \{ i/2\phi (C^{\prime })\} .
\end{equation}
Due to the relations (\ref{5.2}), (\ref{5.9}) all terms 
$|C|^{- 1}{\bf u}^{C}\rho (C)$ in the left hand side sum (\ref{5.1}) 
corresponding with the reduced closed paths $C$ containing the oppositely 
oriented edges ${\bf e}$ and ${\bf e}^{- 1}$ cancel each other. The theorem 
is proved.

Theorem 5.1 is valid also for any graph $G$ embedded in a rectangular 
lattice on the plane. Hence it is possible to change the summing over 
the reduced closed paths on the lattice ${\bf Z}^{\times 2}$ in the 
equality (\ref{4.28}) for the summing over the completely reduced closed 
paths on the lattice ${\bf Z}^{\times 2}$.

The definitions (\ref{2.17}) and (\ref{4.5}) imply
\begin{equation}
\label{5.10}
\rho (({\bf e}_{k},{\bf e}_{1},...,{\bf e}_{k - 1})) =
\rho (({\bf e}_{1},...,{\bf e}_{k})).
\end{equation}
The number $\rho (({\bf e}_{1},...,{\bf e}_{k})) = 
\exp \{ i/2\phi (({\bf e}_{1},...,{\bf e}_{k}))\} $ where 
$\phi (({\bf e}_{1},...,{\bf e}_{k}))$ is the total angle through which 
the tangent vector of the path $({\bf e}_{1},...,{\bf e}_{k})$ turns
along the path $({\bf e}_{1},...,{\bf e}_{k})$.  Therefore 
$\phi (({\bf e}_{1},...,{\bf e}_{k})) = 2\pi m$ where $m$ is an integer.
Hence $(\rho (({\bf e}_{1},...,{\bf e}_{k})))^{- 1} = $

\noindent
$\rho (({\bf e}_{1},...,{\bf e}_{k}))$. The definitions (\ref{2.17}), 
(\ref{4.5}) and the relations (\ref{5.7}) imply
\begin{equation}
\label{5.11}
\rho (({\bf e}_{k}^{- 1},{\bf e}_{k - 1}^{- 1},...,{\bf e}_{1}^{- 1})) =
\rho (({\bf e}_{1},...,{\bf e}_{k})).
\end{equation}
Due to the definitions (\ref{2.18}), (\ref{4.4})
\begin{equation}
\label{5.12}
{\bf u}^{(e_{k}^{- 1},e_{k - 1}^{- 1},...,e_{1}^{- 1})} =
{\bf u}^{(e_{1},...,e_{k})}
\end{equation}
for the reduced closed paths 
$({\bf e}_{k}^{- 1},{\bf e}_{k - 1}^{- 1},...,{\bf e}_{1}^{- 1})$ and
$({\bf e}_{1},...,{\bf e}_{k})$. It follows from the equalities 
(\ref{2.24}), (\ref{5.10}) that the numbers $|C|^{- 1}{\bf u}^{C}\rho (C)$
are equal to each other for $k$ reduced closed paths: 
$({\bf e}_{1},...,{\bf e}_{k})$, 
$({\bf e}_{k},{\bf e}_{1},...,{\bf e}_{k - 1})$,...,
$({\bf e}_{2},...,{\bf e}_{k},{\bf e}_{1})$ which form the oriented 
reduced cycle $[({\bf e}_{1},...,{\bf e}_{k})]$. If the closed path
$({\bf e}_{1},...,{\bf e}_{k})$ is completely reduced, the oriented
cycle $[({\bf e}_{1},...,{\bf e}_{k})]$ is called completely reduced.
The equalities (\ref{5.11}), (\ref{5.12}) imply that the numbers
$|C|^{- 1}{\bf u}^{C}\rho (C)$ are equal to each other for two oriented
reduced cycles $[({\bf e}_{1},...,{\bf e}_{k})]$ and
$[({\bf e}_{k}^{- 1},{\bf e}_{k - 1}^{- 1},...,{\bf e}_{1}^{- 1})]$.
This pair of the oriented reduced cycles $[({\bf e}_{1},...,{\bf e}_{k})]$
and 

\noindent
$[({\bf e}_{k}^{- 1},{\bf e}_{k - 1}^{- 1},...,{\bf e}_{1}^{- 1})]$
is called the non -- oriented reduced cycle 
$[[({\bf e}_{1},...,{\bf e}_{k})]]$. If the closed path
$({\bf e}_{1},...,{\bf e}_{k})$ is completely reduced, the non -- oriented
cycle $[[({\bf e}_{1},...,{\bf e}_{k})]]$ is called completely reduced.
In other words by a non -- oriented completely reduced cycle is meant 
a definite sequence of oriented edges. There are no the oppositely
oriented edges ${\bf e}$, ${\bf e}^{- 1}$ in this sequence. Each 
succeding edge starts at the vertex where the previous edge ended. 
The last edge must end at the vertex from which the first edge started.
The direction in which the sequence of edges is traversed, and also the
particular starting point are both immaterial. By a primitive 
non -- oriented completely reduced cycle is meant one which can not 
be constructed by exactly repeating some subpath of 
$({\bf e}_{1},...,{\bf e}_{k})$ two or more times.

A completely reduced closed path does not contain the oppositely
oriented edges ${\bf e}$, ${\bf e}^{- 1}$. But it may contain the
oriented edge ${\bf e}$ many times. Due to Lemma 2.1 such path is
homotopic to the path $C = ({\bf e},{\bf e}_{1},...,{\bf e}_{k},{\bf e},
{\bf e}_{k + 1},...,{\bf e}_{k + l}) = C_{1}\cdot C_{2}$ where the
closed paths $C_{1} = ({\bf e},{\bf e}_{1},...,{\bf e}_{k})$
and $C_{2} = ({\bf e},{\bf e}_{k + 1},...,{\bf e}_{k + l})$ are
completely reduced. In view of the equality (\ref{2.15})
$\rho (C) = \rho (C_{1})\rho (C_{2})$. A prime closed path is meant a
completely reduced closed path which contains any of its oriented edge 
only once. A non -- oriented completely reduced cycle is called a prime
non -- oriented cycle, if every its representative is a prime closed path.
The prime non -- oriented cycles $[[C_{1}]]$,...,$[[C_{k}]]$ are called
disjoint if any its representatives $C_{1}$,...,$C_{k}$ have no common
oriented edges.

\noindent {\bf Theorem 5.2.} {\it Let a graph} $G$ {\it be embedded in the
rectangular lattice} $\tilde{G} (M_{1}^{\prime },M_{2}^{\prime };M_{1},M_{2})$ 
{\it on the torus. Let with any reduced closed path} $C$ {\it on a graph}
$G$ {\it there correspond the number} $\rho (C)$ {\it given by the 
relations} (\ref{2.17}), (\ref{4.5}) {\it and the number} ${\bf u}^{C}$ 
{\it given by the relations} (\ref{2.18}), (\ref{4.4}). {\it If the estimate}
(\ref{4.6}) {\it is valid, then}
\begin{equation}
\label{5.13}
\exp \{ - 1/2 \sum_{C \in RC(G)} |C|^{- 1}{\bf u}^{C}\rho (C)\} =
1 + \sum_{k = 1}^{\infty }
\sum_{{[[C_{i}]], \, i = 1,...,k :} \atop {prime, \, \, disjoint}}
(- 1)^{k} \bigl( \prod_{i = 1}^{k} {\bf u}^{C_{i}}\bigr)
\bigl( \prod_{i = 1}^{k} \rho (C_{i})\bigr)
\end{equation}
{\it where} $C$ {\it runs over the set} $RC(G)$ {\it of reduced closed
paths on the graph} $G$, $[[C_{i}]]$, $i = 1,...,k$, {\it run over the
set of prime non -- oriented cycles and the prime non -- oriented cycles} 
$[[C_{1}]]$,...,$[[C_{k}]]$ {\it are disjoint.}

\noindent {\it Proof.} Due to Theorem 5.1 it is possible to change the
summing over the reduced closed paths on the graph $G$ in the left hand
side of the equality (\ref{5.13}) for the summing over the completely
reduced closed paths on the graph $G$.

Every completely reduced closed path has the form $C^{\times k}$ where
$k$ is an integer and $C$ is a primitive completely reduced closed path.
By using the equality (\ref{2.22}) we get
\begin{equation}
\label{5.14}
\exp \{ - 1/2\sum_{C \in RC(G)} |C|^{- 1}{\bf u}^{C}\rho (C) \} =
\exp \{ 1/2\sum_{{C \in CRC(G),} \atop primitive} 
|C|^{- 1}\ln (1 - {\bf u}^{C}\rho (C)) \}.
\end{equation}
Let a closed path $({\bf e}_{1},...,{\bf e}_{k})$ be a primitive
completely reduced one. Then $k$ primitive completely reduced closed 
paths $({\bf e}_{1},...,{\bf e}_{k})$, 
$({\bf e}_{2},...,{\bf e}_{k},{\bf e}_{1})$,...,
$({\bf e}_{k},{\bf e}_{1},...,{\bf e}_{k - 1})$ are different. They form
a primitive oriented completely reduced cycle 
$[({\bf e}_{1},...,{\bf e}_{k})]$. Due to the equalities (\ref{2.24}),
(\ref{5.10}) the numbers are equal to each other for all representatives
of this primitive oriented completely reduced cycle
$[({\bf e}_{1},...,{\bf e}_{k})]$. Hence
\begin{equation}
\label{5.15}
\exp \{ - 1/2\sum_{C \in RC(G)} |C|^{- 1}{\bf u}^{C}\rho (C) \} =
\exp \{ 1/2\sum_{{[C]: \, \, C \in CRC(G),} \atop primitive} 
\ln (1 - {\bf u}^{C}\rho (C)) \}.
\end{equation}
Due to the relations (\ref{5.11}), (\ref{5.12}) the numbers
${\bf u}^{C}\rho (C)$ are equal to each other for all representatives
of the primitive non -- oriented completely reduced cycle
$[[({\bf e}_{1},...,{\bf e}_{k})]]$. We change the summing over the set 
of primitive oriented completely reduced cycle in the right hand side
of the equality (\ref{5.15}) for the summing over the set of primitive
non -- oriented completely reduced cycles. Thus the right hand side of 
the equality (\ref{5.15}) is the product of the multipliers
$(1 - {\bf u}^{C}\rho (C))$. The decomposition of this product into the
series gives
\begin{eqnarray}
\label{5.16}
\exp \{ - 1/2 \sum_{C \in RC(G)} |C|^{- 1}{\bf u}^{C}\rho (C)\} =
\nonumber \\
1 + \sum_{k = 1}^{\infty }
\sum_{{[[C_{i}]],\, \, C_{i} \in CRC(G),\, \, i = 1,...,k:}
\atop {primitive, \, \, different}}
(- 1)^{k} \bigl( \prod_{i = 1}^{k} {\bf u}^{C_{i}}\bigr)
\bigl( \prod_{i = 1}^{k} \rho (C_{i})\bigr)
\end{eqnarray}
where $[[C_{i}]]$, $i = 1,...,k$, run over the set of primitive 
non -- oriented completely reduced cycles and the primitive 
non -- oriented completely reduced cycles $[[C_{1}]]$,...,$[[C_{k}]]$
differ from each other.

Let us choose an oriented edge ${\bf e}$. Any completely reduced closed 
path does not contain the oppositely oriented edges ${\bf e}$ and 
${\bf e}^{- 1}$ simultaneously. We choose the representatives $C_{i}$
of the non -- oriented completely reduced cycles $[[C_{i}]]$ in the
right hand side of the equality (\ref{5.16}) such that the paths $C_{i}$
do not contain the oriented edge ${\bf e}^{- 1}$. Thus the sum of all
terms in the series (\ref{5.16}) which contains $u({\bf e})^{n}$ is
proportional to the sum
\begin{equation}
\label{5.17}
\sum_{k = 1}^{n}
\sum_{{[C_{i}],\, \, C_{i} \in CRC(G),\, \, i = 1,...,k:}
\atop {primitive, \, \, different}}
(- 1)^{k} \bigl( \prod_{i = 1}^{k} {\bf u}^{C_{i}}\bigr)
\bigl( \prod_{i = 1}^{k} \rho (C_{i})\bigr)
\end{equation}
where the primitive completely reduced closed path $C_{i}$ contains
(it may be many times) the oriented edge ${\bf e}$. The different paths
$C_{1}$,...,$C_{k}$ all together contain the oriented edge ${\bf e}$
exactly $n$ times. Due to Lemma 2.1 we can choose such representatives
$C_{1}$,...,$C_{k}$ of the primitive oriented completely reduced cycles
$[C_{1}]$,...,$[C_{k}]$ that all paths start with the oriented edge 
${\bf e}$. Every such path $C_{j}$ is the product 
$C_{j1}^{\prime }\cdots C_{jq_{j}}^{\prime }$ where the completely reduced 
closed path $C_{jl}^{\prime }$ starts with the oriented edge ${\bf e}$
and contains it exactly one time. All paths $C_{jl}^{\prime }$ are
primitive. Since the paths $C_{1}$,...,$C_{k}$ all together contain
the oriented edge ${\bf e}$ exactly $n$ times, they are decomposed into
$n$ paths $C_{jl}^{\prime }$. We change the numeration so that
$C_{jl}^{\prime } = C_{i_{jl}}^{\prime }$ where the numbers
$(i_{j1},...,i_{jq_{j}})$, $j = 1,...,k$, give the subdivision of the
numbers $1,...,n$ into $k$ groups. Due to definitions (\ref{2.17}),
(\ref{4.5}) the number $\rho (C_{j}) = \rho (C_{i_{j1}}^{\prime })\cdots 
\rho (C_{i_{jq_{j}}}^{\prime })$. The definitions (\ref{2.18}), (\ref{4.4})
imply that ${\bf u}^{C_{j}} = {\bf u}^{C_{i_{j1}}^{\prime }}\cdots 
{\bf u}^{C_{i_{jq_{j}}}^{\prime }}$. Thus the sum (\ref{5.17}) has the
following form
\begin{equation}
\label{5.18}
\sum_{{[C_{i}^{\prime }],\, \, C_{i}^{\prime } \in CRC(G),\, \, i = 1,...,n:}
\atop {primitive}}
\bigl( \prod_{i = 1}^{n} {\bf u}^{C_{i}^{\prime }}\bigr)
\bigl( \prod_{i = 1}^{n} \rho (C_{i}^{\prime })\bigr)
\sum_{k = 1}^{n}
\sum_{{[C_{i_{j1}}^{\prime}\cdots C_{i_{jq_{j}}}^{\prime}],\, \, j = 1,...,k:}
\atop {primitive,\, \, different}} (- 1)^{k}
\end{equation}
where the primitive completely reduced closed paths $C_{1}^{\prime }$,
...,$C_{n}^{\prime }$ start with the oriented edge ${\bf e}$ and
contain it exactly one time. The numbers $(i_{j1},...,i_{jq_{j}})$,
$j = 1,...,k$, give a subdivision of the numbers $1,...,n$ into $k$
groups.

If the sum (\ref{5.18}) is equal to zero for $n > 1$, then the left hand
sides of the equalities (\ref{5.13}) and (\ref{5.16}) coincide and the
theorem is proved.

Let us consider first the case when all completely reduced closed paths
$C_{1}^{\prime }$,...,$C_{n}^{\prime }$ are different. With every oriented
completely reduced cycle 
$[C_{i_{j1}}^{\prime}\cdots C_{i_{jq_{j}}}^{\prime}]$ there corresponds
the permutation
$\pi [i_{j1},...,i_{jq_{j}}]: i_{jl} \rightarrow i_{j,l + 1}$,
$l = 1,...,q_{j} - 1$, $i_{jq_{j}} \rightarrow i_{j1}$. All the other
numbers from $1,...,n$ the permutation $\pi [i_{j1},...,i_{jq_{j}}]$
leaves invariant. Since all completely reduced closed paths $C_{1}^{\prime }$,
...,$C_{n}^{\prime }$ are different, any permutation of the 
multipliers in the product 
$C_{i_{j1}}^{\prime}\cdots C_{i_{jq_{j}}}^{\prime}$ gives another 
completely reduced closed path. The permutation $\pi [i_{j1},...,i_{jq_{j}}]$
gives the completely reduced closed path $C_{i_{j2}}^{\prime }\cdots 
C_{i_{jq_{j}}}^{\prime }\cdot C_{i_{j1}}^{\prime }$.
Due to Lemma 2.1 it is homotopic to the completely reduced closed path
$C_{i_{j1}}^{\prime}\cdots C_{i_{jq_{j}}}^{\prime}$. But the permutations
$\pi [i_{j2},...,i_{jq_{j}},i_{j1}]$ and $\pi [i_{j1},...,i_{jq_{j}}]$
coincide. Hence the correspondence between the oriented completely reduced 
cycles $[C_{i_{j1}}^{\prime}\cdots C_{i_{jq_{j}}}^{\prime}]$ and the
permutations $\pi [i_{j1},...,i_{jq_{j}}]$ is one -- to -- one. If the
groups of numbers $i_{j1}$,...,$i_{jq_{j}}$ and $i_{j^{\prime }1}$,...,
$i_{j^{\prime }q_{j^{\prime }}}$ do not intersect, then the permutations
$\pi [i_{j1},...,i_{jq_{j}}]$ and 
$\pi [i_{j^{\prime }1},...,i_{j^{\prime }q_{j^{\prime }}}]$ commute with
each other. Thus any subdivision of the numbers $1,...,n$ into $k$
non -- intersecting groups $i_{j1},...,i_{jq_{j}}$, $j = 1,...,k$
corresponds with the set of $k$ oriented completely reduced cycles
$[C_{i_{j1}}^{\prime}\cdots C_{i_{jq_{j}}}^{\prime}]$, $j = 1,...,k$,
and with the permutation
$\pi [i_{11},...,i_{1q_{1}}]\cdots \pi [i_{k1},...,i_{kq_{k}}]$ of
the numbers $1,...,n$. Conversely, any permutation $\pi $ from the
permutation group $S_{n}$ of the numbers $1,...,n$ corresponds with
a subdivision of the numbers $1,...,n$ into the systems of transitivity
of the permutation 
$\pi $: $(i_{j1},\pi (i_{j1}),...,\pi^{q_{j} - 1} (i_{j1}))$ where 
$\pi^{q_{j}} (i_{j1}) = i_{j1}$. The total number of these systems of 
transitivity is denoted by $t(\pi )$. Any system of transitivity of the
permutation $\pi $ corresponds with an oriented completely reduced 
cycle $[C_{i_{j1}}^{\prime }\cdot C_{\pi (i_{j1})}^{\prime }\cdots 
C_{\pi^{q_{j} - 1} (i_{j1})}^{\prime }]$. Therefore for the different 
completely reduced closed paths $C_{1}^{\prime }$,...,$C_{n}^{\prime }$ 
the following relation is valid
\begin{equation}
\label{5.19}
\sum_{k = 1}^{n}
\sum_{[C_{i_{j1}}^{\prime}\cdots C_{i_{jq_{j}}}^{\prime}],\, \, j = 1,...,k}
(- 1)^{k} = \sum_{\pi \in S_{n}} (- 1)^{t(\pi )}.
\end{equation}
Since all completely reduced closed paths $C_{1}^{\prime }$,...,
$C_{n}^{\prime }$ are different, all oriented completely reduced
cycles $[C_{i_{j1}}^{\prime}\cdots C_{i_{jq_{j}}}^{\prime}]$, 
$j = 1,...,k$, are primitive and different.

Let us define $n\times n$ -- matrix $A$ whose matrix elements 
$A_{ij} = - 1$. We calculate the determinant of this matrix 
\begin{equation}
\label{5.20}
\det A = \sum_{\pi \in S_{n}} (- 1)^{\sigma (\pi )}
\prod_{i = 1}^{t(\pi )} A_{p_{i}\pi (p_{i})}A_{\pi (p_{i})\pi^{2} (p_{i})}
\cdots A_{\pi^{q_{i} - 1} (p_{i})p_{i}}
\end{equation}
where $\pi^{q_{i}} (p_{i}) = p_{i}$ and the numbers
$(p_{i},\pi (p_{i}),...,\pi^{q_{i} - 1} (p_{i}))$, $i = 1,...,t(\pi )$,
give the subdivision of the numbers $1,...,n$ into $t(\pi )$ groups.
The parity of the permutation $\pi $ is equal to
\begin{equation}
\label{5.21}
\sigma (\pi ) = \sum_{i = 1}^{t(\pi )} (q_{i} - 1) \, \, \hbox{mod} \, 2.
\end{equation}
The substitution of the relation (\ref{5.21}) and of the matrix elements
$A_{ij} = - 1$ into the relation (\ref{5.20}) gives
\begin{equation}
\label{5.22}
\det A = \sum_{\pi \in S_{n}} (- 1)^{t(\pi )}.
\end{equation}
$\det A = 0$ for $n > 1$. Hence the sum (\ref{5.19}) equals zero for
$n > 1$.

If all completely reduced closed paths 
$C_{1}^{\prime }$,...,$C_{n}^{\prime }$ coincide, then the completely 
reduced closed path $C_{i_{j1}}^{\prime}\cdots C_{i_{jq_{j}}}^{\prime}$ 
is non -- primitive for $q_{j} > 1$. If all $q_{j} = 1$, the completely 
reduced closed paths $C_{1}^{\prime }$,...,$C_{n}^{\prime }$ are 
not different. Thus for this case the sum (\ref{5.18}) does not give the 
contribution into the sum (\ref{5.16}).

Let us consider the case when there are $m$ groups in which $n_{i} > 1$,
$i = 1,...,m$, the completely reduced closed paths $C_{j}^{\prime }$
coincide with each other. We prove the following equality
\begin{eqnarray}
\label{5.23}
\sum_{k = 1}^{n}
\sum_{{[C_{i_{j1}}^{\prime}\cdots C_{i_{jq_{j}}}^{\prime}],\, \, j = 1,...,k:}
\atop {primitive,\, \, different}} (- 1)^{k} = \nonumber \\
\bigl( \prod_{i = 1}^{m} (n_{i})!\bigr)^{- 1}
\bigl[ \sum_{\pi \in S_{n}} (- 1)^{t(\pi )} -
\sum_{l > 1, \, \, q \geq 1}^{\prime } \sum_{\pi \in S_{n}}^{\prime}
(- 1)^{t(\pi ) - lq} \sum_{\tau \in S_{l}} (- 1)^{t(\tau )} \bigr] .
\end{eqnarray}
In the right hand side of the equality (\ref{5.23}) the second sum
extends over certain numbers $l,q$ and over certain permutations
$\pi \in S_{n}$.
 
Now the right hand side sum (\ref{5.19}) contains the permutations 
corresponding with the non -- primitive and non -- different oriented 
completely reduced cycles. Let the completely reduced closed paths 
$C_{i_{1j}}^{\prime }$,$C_{i_{2j}}^{\prime }$,...,$C_{i_{lj}}^{\prime }$, 
$j = 1,...,q$, coincide with each other. Let $\tau \in S_{l}$ be a 
permutation of the numbers $1,...,l$. The groups of the numbers 
$(p_{j},\tau (p_{j}),...,\tau^{d_{j} - 1} (p_{j}))$, 
$\tau^{d_{j}} (p_{j}) = p_{j}$, $j = 1,...,t(\tau )$, give the subdivision
of the numbers $1,...,l$ into the systems of transitivity of the
permutation $\tau $. We consider the permutations

\noindent $\pi \{ \tau ,j\} \equiv
\pi [i_{p_{j}1},...,i_{p_{j}q},i_{\tau (p_{j})1},...,i_{\tau (p_{j})q},
i_{\tau^{d_{j} - 1} (p_{j})1},...,i_{\tau^{d_{j} - 1} (p_{j})q}]$ of the 
numbers $1,...,n$ where $j = 1,...,t(\tau )$. The permutations 
$\pi \{ \tau ,j_{1}\} $ and $\pi \{ \tau ,j_{2}\} $ commute with each
other. Let $\pi \{ \tau \} = \pi \{ \tau ,1\} \cdots \pi \{ \tau ,t(\tau )\}$.
The permutation leaves invariant $n - ql$ numbers from $1,...,n$. 
By the construction $t(\pi \{ \tau \} ) = t(\tau ) + n - ql$. Let
a permutation $\pi^{\prime } \in S_{n}$ leave invariant the numbers
$i_{1j}$,...,$i_{lj}$, $j = 1,...,q$. Then
\begin{equation}
\label{5.24}
t(\pi^{\prime } \cdot \pi \{ \tau \} ) = t(\pi^{\prime } ) - lq + t(\tau ).
\end{equation}
The permutation $\pi^{\prime } \cdot \pi \{ \tau \} \in S_{n}$ 
corresponds with $t(\pi^{\prime } ) - lq + t(\tau )$ oriented
completely reduced cycles. From these cycles $t(\tau )$ oriented
completely reduced cycles $[C_{i_{p_{j}1}}^{\prime }\cdots 
C_{i_{p_{j}q}}^{\prime }\cdot C_{i_{\tau (p_{j})1}}^{\prime }\cdots 
C_{i_{\tau (p_{j})q}}^{\prime }\cdots 
C_{i_{\tau^{d_{j} - 1} (p_{j})1}}^{\prime }\cdots
C_{i_{\tau^{d_{j} - 1} (p_{j})q}}^{\prime }]$, $j = 1,...,t(\tau )$,
are non -- primitive for $d_{j} > 1$. If all numbers $d_{j} = 1$
and $t(\tau ) = l$, then these $t(\tau )$ oriented completely
reduced cycles are non -- different. Hence the permutation 
$\pi^{\prime } \cdot \pi \{ \tau \} $ must be subtracted from the 
right hand side sum (\ref{5.19}). Now we explain the multiplier
$(\prod_{i = 1}^{m} (n_{i})!)^{- 1}$ in the right hand side of the
equality (\ref{5.23}).

Let a permutation $\lambda \in S_{n}$ rearrange only the numbers
corresponding with the coinciding completely reduced closed paths
$C_{1}^{\prime }$,...,$C_{n}^{\prime }$. For this permutation
$C_{\lambda (i_{j1})}^{\prime }\cdots C_{\lambda (i_{jq_{j}})}^{\prime } =
C_{i_{j1}}^{\prime }\cdots C_{i_{jq_{j}}}^{\prime }$. But the
permutations $\pi [\lambda (i_{j1}),...,\lambda (i_{jq_{j}})]$ and
$\pi [i_{j1},...,i_{jq_{j}}]$ may coincide only in the case when the
permutation $\lambda $ acts on the numbers $i_{j1}$,...,$i_{jq_{j}}$
as a cyclic permutation. By the definition of the permutation $\lambda $
it is possible only for the coinciding completely reduced closed paths
$C_{i_{j1}}^{\prime }$,..., $C_{i_{jq_{j}}}^{\prime }$. But then the 
oriented completely reduced cycle
$[C_{i_{j1}}^{\prime }\cdots C_{i_{jq_{j}}}^{\prime }]$ is non -- primitive.
Thus every set of the different primitive oriented completely reduced
cycles $[C_{i_{j1}}^{\prime }\cdots C_{i_{jq_{j}}}^{\prime }]$,
$j = 1,...,k$, corresponds with $\prod_{i = 1}^{m} (n_{i})! $
different permutations $\pi [\lambda (i_{11}),...,\lambda (i_{1q_{1}})]
\cdots \pi [\lambda (i_{k1}),...,\lambda (i_{kq_{k}})]$ of the numbers
$1,...,n$. The number $\prod_{i = 1}^{m} (n_{i})! $ is the total number of
the permutations $\lambda $ rearranging only the numbers corresponding
with the coinciding completely reduced closed paths 
$C_{1}^{\prime }$,...,$C_{n}^{\prime }$. The equality (\ref{5.23}) is proved.

The equalities (\ref{5.22}), (\ref{5.23}) imply that the left hand side sum 
(\ref{5.23}) is equal to zero for $n > 1$. Hence the sum (\ref{5.18})
equals zero for $n > 1$ and the right hand sides of the equalities
(\ref{5.13}) and (\ref{5.16}) coincide. The theorem is proved.

For the proof of Theorem 5.2 we used Theorem 5.1 and the definitions
(\ref{2.17}), (\ref{4.5}), and (\ref{2.18}), (\ref{4.4}). Therefore
Theorem 5.2 is valid also for a graph $G$ embedded in a rectangular
lattice on the plane.

Let the left hand side of the inequality (\ref{2.30}) be denoted by
$||u||$. The inequality (\ref{4.6}) is a particular case of the 
inequality (\ref{2.30}). It may be rewritten as $||u|| < 1/3$.

\noindent {\bf Theorem 5.3.} {\it Let a graph} $G$ {\it be embedded in
the rectangular lattice} 
$\tilde{G} (M_{1}^{\prime },M_{2}^{\prime };M_{1},M_{2})$ {\it on the
torus. Let the estimate} (\ref{4.6}) {\it be fulfilled and let
interaction energy} $E({\bf e})$ {\it be non -- negative. Then for the
reduced partition function} (\ref{4.2}) {\it the following inequalities
are valid}
\begin{eqnarray}
\label{5.25}
1 - 8/3 \bigl( \prod_{s = 1}^{2} (M_{s} - M_{s}^{\prime })\bigr)
(1 - 3||u||)^{- 1}\sum_{s = 1}^{2} (3||u||)^{M_{s} - M_{s}^{\prime }} \leq
\nonumber \\
(Z_{r,G})^{- 1}
\exp \{ - 1/2\sum_{C \in RC(G)} |C|^{- 1}{\bf u}^{C}\rho (C)\} \leq 1
\end{eqnarray}
{\it where} $C$ {\it run over the set} $RC(G)$ {\it of reduced closed
paths on the graph} $G$, {\it the natural number} $|C|$ {\it is the
length of the path} $C$, {\it the number} ${\bf u}^{C}$ {\it is given
by the equalities} (\ref{2.18}), (\ref{4.4}) {\it and the number}
$\rho (C)$ {\it is given by the equalities} (\ref{2.17}), (\ref{4.5}).

\noindent {\it Proof.} Due to Theorem 5.2 the equality (\ref{5.13})
is valid. Let four different oriented edges ${\bf e}_{i}$, $i = 1,...,4$,
of the graph $G$ have the same initial vertex: 
$b({\bf e}_{1}) = \cdots = b({\bf e}_{4})$. Let the prime closed paths
$C_{1} = ({\bf e}_{1},{\bf e}_{5},...,{\bf e}_{m},{\bf e}_{2}^{- 1})$ and
$C_{2} = ({\bf e}_{3},{\bf e}_{m + 1},...,{\bf e}_{m + n},{\bf e}_{4}^{- 1})$
on the graph $G$ correspond with the disjoint prime non -- oriented
cycles $[[C_{1}]]$ and $[[C_{2}]]$. Then the products $C_{1}\cdot C_{2} =
({\bf e}_{1},{\bf e}_{5},...,{\bf e}_{m},{\bf e}_{2}^{- 1},{\bf e}_{3},
{\bf e}_{m + 1},...,{\bf e}_{m + n},{\bf e}_{4}^{- 1})$ and

\noindent
$C_{1}\cdot C_{2}^{- 1} = ({\bf e}_{1},{\bf e}_{5},...,{\bf e}_{m},
{\bf e}_{2}^{- 1},{\bf e}_{4},{\bf e}_{m + n}^{- 1},...,{\bf e}_{m + 1}^{- 1},
{\bf e}_{3}^{- 1})$ are the prime closed paths. Let $C_{3}$,...,$C_{k}$ be
the prime closed paths on the graph $G$ such that the non -- oriented cycles

\noindent
$[[C_{1}]]$,$[[C_{2}]]$,$[[C_{3}]]$,...,$[[C_{k}]]$ are disjoint.
Then the prime non -- oriented cycles $[[C_{1}\cdot C_{2}]]$,
$[[C_{3}]]$,...,$[[C_{k}]]$ are disjoint and the prime non -- oriented
cycles $[[C_{1}\cdot C_{2}^{- 1}]]$,$[[C_{3}]]$,...,$[[C_{k}]]$ are
disjoint too.

Let us consider the sum of three terms in the right hand side sum 
(\ref{5.13})
\begin{eqnarray}
\label{5.26}
(- 1)^{k}\bigl( \prod_{i = 1}^{k} {\bf u}^{C_{i}}\bigr)
\bigl( \prod_{i = 1}^{k} \rho (C_{i})\bigr) +
(- 1)^{k - 1}{\bf u}^{C_{1}\cdot C_{2}} 
\bigl( \prod_{i = 3}^{k} {\bf u}^{C_{i}}\bigr)
\rho (C_{1}\cdot C_{2})\bigl( \prod_{i = 3}^{k} \rho (C_{i})\bigr)  +
\nonumber \\
(- 1)^{k - 1}{\bf u}^{C_{1}\cdot C_{2}^{- 1}} 
\bigl( \prod_{i = 3}^{k} {\bf u}^{C_{i}}\bigr)
\rho (C_{1}\cdot C_{2}^{- 1})\bigl( \prod_{i = 3}^{k} \rho (C_{i})\bigr).
\end{eqnarray}
The definitions (\ref{2.18}), (\ref{4.4}) imply that
\begin{equation}
\label{5.27}
{\bf u}^{C_{1}\cdot C_{2}} = {\bf u}^{C_{1}\cdot C_{2}^{- 1}} =
{\bf u}^{C_{1}}{\bf u}^{C_{2}}.
\end{equation}
The definitions (\ref{2.17}), (\ref{4.5}) and the relations (\ref{5.7})
imply
\begin{eqnarray}
\label{5.28}
\rho (C_{1}\cdot C_{2}) = 
\gamma ({\bf e}_{1},{\bf e}_{2},{\bf e}_{3},{\bf e}_{4})\rho (C_{1})
\rho (C_{2}), \nonumber \\
\rho (C_{1}\cdot C_{2}^{- 1}) = 
\gamma ({\bf e}_{1},{\bf e}_{2},{\bf e}_{4},{\bf e}_{3})\rho (C_{1})
\rho (C_{2})
\end{eqnarray}
where
\begin{equation}
\label{5.29}
\gamma ({\bf e}_{1},{\bf e}_{2},{\bf e}_{3},{\bf e}_{4})  =
\rho ({\bf e}_{1}^{- 1};{\bf e}_{2})\rho ({\bf e}_{2}^{- 1};{\bf e}_{3})
\rho ({\bf e}_{3}^{- 1};{\bf e}_{4})\rho ({\bf e}_{4}^{- 1};{\bf e}_{1}).
\end{equation}
In the second equality (\ref{5.28}) we took into account the value of the
angle through which the tangent vector of the path $C_{2}$ turns
along the path $C_{2}$: $\phi (C_{2}) = 2\pi k$ where $k$ is an integer.
Hence $\rho (C_{2}) = \exp \{ i/2\phi (C_{2})\} = (\rho (C_{2}))^{- 1}$.

The substitution of the equalities (\ref{5.27}), (\ref{5.28}) into the
expression (\ref{5.26}) gives
\begin{equation}
\label{5.30}
[1 - \gamma ({\bf e}_{1},{\bf e}_{2},{\bf e}_{3},{\bf e}_{4}) -
\gamma ({\bf e}_{1},{\bf e}_{2},{\bf e}_{4},{\bf e}_{3})]
(- 1)^{k}\bigl( \prod_{i = 1}^{k} {\bf u}^{C_{i}}\bigr)
\bigl( \prod_{i = 1}^{k} \rho (C_{i})\bigr) .
\end{equation}

The definition (\ref{5.29}) implies
\begin{equation}
\label{5.31}
\gamma ({\bf e}_{2},{\bf e}_{3},{\bf e}_{4},{\bf e}_{1}) =
\gamma ({\bf e}_{1},{\bf e}_{2},{\bf e}_{3},{\bf e}_{4}).
\end{equation}
Let for some index $i = 1,...,3$ the oriented edges ${\bf e}_{i}$ and
${\bf e}_{i + 1}$ have the opposite directions or let the oriented
edges ${\bf e}_{1}$ and ${\bf e}_{4}$ have the opposite directions.
We shall prove that for all these cases 
$\gamma ({\bf e}_{1},{\bf e}_{2},{\bf e}_{3},{\bf e}_{4}) = 1$. Due to
the relation (\ref{5.31}) it is sufficient to prove this statement only
for the case when the oriented edges ${\bf e}_{1}$ and ${\bf e}_{2}$
have the opposite directions. In this case the oriented edges ${\bf e}_{3}$
and ${\bf e}_{4}$ have also the opposite directions. Due to the
definitions (\ref{4.5}), (\ref{5.29}) we have in this case
\begin{equation}
\label{5.32}
\gamma ({\bf e}_{1},{\bf e}_{2},{\bf e}_{3},{\bf e}_{4}) =
\rho ({\bf e}_{2}^{- 1};{\bf e}_{3})\rho ({\bf e}_{4}^{- 1};{\bf e}_{1}).
\end{equation}
The direction of the oriented edge ${\bf e}_{2}^{- 1}$ coincides with the
direction of the oriented edge ${\bf e}_{1}$. The direction of the 
oriented edge ${\bf e}_{4}^{- 1}$ coincides with the direction of the 
oriented edge ${\bf e}_{3}$. Hence the definition (\ref{4.5}) and the
equality (\ref{5.32}) imply
$\gamma ({\bf e}_{1},{\bf e}_{2},{\bf e}_{3},{\bf e}_{4}) = 1$.

It is easy to verify that when the directions of the oriented edges 
${\bf e}_{i}$ and ${\bf e}_{i + 1}$, $i = 1,...,3$, ${\bf e}_{4}$ and
${\bf e}_{1}$ are orthogonal to each other the definitions
(\ref{4.5}), (\ref{5.29}) imply
$\gamma ({\bf e}_{1},{\bf e}_{2},{\bf e}_{3},{\bf e}_{4}) = - 1$.

Let the directions of the oriented edges ${\bf e}_{i}$
and ${\bf e}_{i + 1}$, $i = 1,...,3$, ${\bf e}_{4}$ and ${\bf e}_{1}$
be orthogonal to each other. Then the directions of the oriented edges 
${\bf e}_{2}$ and ${\bf e}_{4}$ are opposite. Therefore
$\gamma ({\bf e}_{1},{\bf e}_{2},{\bf e}_{3},{\bf e}_{4}) = - 1$ and
$\gamma ({\bf e}_{1},{\bf e}_{2},{\bf e}_{4},{\bf e}_{3}) = 1$. Hence
due to the relations (\ref{5.27}), (\ref{5.28}) in this case the
expression (\ref{5.30}) is equal to
\begin{equation}
\label{5.33}
(- 1)^{k}\bigl( \prod_{i = 1}^{k} {\bf u}^{C_{i}}\bigr)
\bigl( \prod_{i = 1}^{k} \rho (C_{i})\bigr) =
(- 1)^{k - 1}{\bf u}^{C_{1}\cdot C_{2}} 
\bigl( \prod_{i = 3}^{k} {\bf u}^{C_{i}}\bigr)
\rho (C_{1}\cdot C_{2})\bigl( \prod_{i = 3}^{k} \rho (C_{i})\bigr) .  
\end{equation}
The paths $C_{1}$,$C_{2}$ and $C_{1}\cdot C_{2}$ go subsequently through
the oriented edges ${\bf e}_{i}^{\pm 1}$, $i = 1,...,4$, having the
directions orthogonal to each other.

Let the oriented edges ${\bf e}_{1}$ and ${\bf e}_{2}$, ${\bf e}_{2}$
and ${\bf e}_{4}$, ${\bf e}_{4}$ and ${\bf e}_{3}$, ${\bf e}_{3}$ and
${\bf e}_{1}$ have the directions orthogonal to each other. Thus the 
oriented edges ${\bf e}_{2}$ and ${\bf e}_{3}$ have the opposite directions. 
Therefore
$\gamma ({\bf e}_{1},{\bf e}_{2},{\bf e}_{3},{\bf e}_{4}) = 1$ and
$\gamma ({\bf e}_{1},{\bf e}_{2},{\bf e}_{4},{\bf e}_{3}) = - 1$.
Hence in this case the expression (\ref{5.30}) is equal to
\begin{equation}
\label{5.34}
(- 1)^{k}\bigl( \prod_{i = 1}^{k} {\bf u}^{C_{i}}\bigr)
\bigl( \prod_{i = 1}^{k} \rho (C_{i})\bigr) =
(- 1)^{k - 1}{\bf u}^{C_{1}\cdot C_{2}^{- 1}} 
\bigl( \prod_{i = 3}^{k} {\bf u}^{C_{i}}\bigr)
\rho (C_{1}\cdot C_{2}^{- 1})\bigl( \prod_{i = 3}^{k} \rho (C_{i})\bigr).
\end{equation}
The paths $C_{1}$,$C_{2}$ and $C_{1}\cdot C_{2}^{- 1}$ go subsequently
through the oriented edges ${\bf e}_{i}^{\pm 1}$, $i = 1,...,4$,
having the directions orthogonal to each other.

Let the oriented edges ${\bf e}_{1}$ and ${\bf e}_{2}$ have the opposite
directions. Hence
$\gamma ({\bf e}_{1},{\bf e}_{2},{\bf e}_{3},{\bf e}_{4}) =
\gamma ({\bf e}_{1},{\bf e}_{2},{\bf e}_{4},{\bf e}_{3}) = 1$ and 
in this case the expression (\ref{5.30}) is equal to
\begin{eqnarray}
\label{5.35}
(- 1)^{k - 1}{\bf u}^{C_{1}\cdot C_{2}} 
\bigl( \prod_{i = 3}^{k} {\bf u}^{C_{i}}\bigr)
\rho (C_{1}\cdot C_{2})\bigl( \prod_{i = 3}^{k} \rho (C_{i})\bigr)  =
\nonumber \\
(- 1)^{k - 1}{\bf u}^{C_{1}\cdot C_{2}^{- 1}} 
\bigl( \prod_{i = 3}^{k} {\bf u}^{C_{i}}\bigr)
\rho (C_{1}\cdot C_{2}^{- 1})\bigl( \prod_{i = 3}^{k} \rho (C_{i})\bigr).
\end{eqnarray}
Since the oriented edges ${\bf e}_{1}$ and ${\bf e}_{2}$ have the
opposite directions, the oriented edges ${\bf e}_{3}$ and
${\bf e}_{4}$ have the opposite directions and the directions of the
oriented edges ${\bf e}_{1}$ and ${\bf e}_{4}$, ${\bf e}_{4}$ and
${\bf e}_{2}$, ${\bf e}_{2}$ and ${\bf e}_{3}$, ${\bf e}_{3}$ and
${\bf e}_{1}$ are orthogonal to each other. Thus in this case the
paths $C_{1}\cdot C_{2}$ and $C_{1}\cdot C_{2}^{- 1}$ go subsequently
through the oriented edges ${\bf e}_{i}^{\pm 1}$, $i = 1,...,4$, having
the directions orthogonal to each other.

We have considered all possible directions of the oriented edges
${\bf e}_{1}$,${\bf e}_{2}$,${\bf e}_{3}$,${\bf e}_{4}$. In any case the
sum (\ref{5.26}) is equal to one of the expressions 
(\ref{5.33}) -- (\ref{5.35}) where the paths go subsequently through
the oriented edges ${\bf e}_{i}^{\pm 1}$, $i = 1,...,4$, having the
directions orthogonal to each other.

With the term
\begin{equation}
\label{5.36}
(- 1)^{k}\bigl( \prod_{i = 1}^{k} {\bf u}^{C_{i}}\bigr)
\bigl( \prod_{i = 1}^{k} \rho (C_{i})\bigr)
\end{equation}
the right hand side sum (\ref{5.13}) contains all the terms of type 
(\ref{5.36}) where the disjoint prime non -- oriented cycles
$[[C_{1}^{\prime }]]$,...,$[[C_{k^{\prime }}^{\prime }]]$ contain the
same non -- oriented edges as the disjoint prime non -- oriented cycles
$[[C_{1}]]$,...,$[[C_{k}]]$. We have proved that the sum of these terms
of type (\ref{5.36}) is equal to the only term of type (\ref{5.36}) where
the prime closed paths $C_{1}$,...,$C_{k}$ satisfy the condition: if four 
different oriented edges from the prime closed paths $C_{1}$,...,$C_{k}$
are incident to one vertex, then the prime closed paths $C_{1}$,...,$C_{k}$
go subsequently through the oriented edges from these four oriented edges
having the directions orthogonal to each other.

Let us define the cochain $\xi^{1} [C] \in C^{1}(P(G),{\bf Z}_{2}^{add})$
equal $1$ on all non -- oriented edges from the prime non -- oriented
cycle $[[C]]$ on the graph $G$. The cochain $\xi^{1} [C]$ equals zero on
all other non -- oriented edges of the graph $G$. Since $[[C]]$ is a prime
non -- oriented cycle, $\xi^{1} [C] \in Z_{1}(P(G),{\bf Z}_{2}^{add})$.
If the prime non -- oriented cycles $[[C_{1}]]$,...,$[[C_{k}]]$ are
disjoint, then the supports of the cycles 
$\xi^{1} [C_{1}]$,...,$\xi^{1} [C_{k}]$ do not intersect. By the definition
$\rho (C_{i}) = \pm 1$, $i = 1,...,k$. The interaction energy $E({\bf e})$
is non -- negative. Hence we obtain the estimate for the term (\ref{5.36})
\begin{equation}
\label{5.37}
(- 1)^{k}\bigl( \prod_{i = 1}^{k} {\bf u}^{C_{i}}\bigr)
\bigl( \prod_{i = 1}^{k} \rho (C_{i})\bigr) \leq
{\bf u}^{\sum_{1}^{k} \xi^{1} [C_{i}]}.
\end{equation}
Since the sum of the terms of type (\ref{5.36}) corresponding with
the disjoint prime non -- oriented cycles containing the same set of the
non -- oriented edges is again the term of type (\ref{5.36}), it
follows from the equality (\ref{4.2}) and from the inequality (\ref{5.37})
that
\begin{equation}
\label{5.38}
\exp \{ - 1/2\sum_{C \in RC(G)} |C|^{- 1}{\bf u}^{C}\rho (C)\} \leq
Z_{r,G}.
\end{equation}

We shall prove that for any cycle $\xi^{1} \in Z_{1}(P(G),{\bf Z}_{2}^{add})$
there exist the disjoint prime non -- oriented cycles
$[[C_{1}]]$,...,$[[C_{k}]]$ such that
$\xi^{1} = \xi^{1} [C_{1}] + \cdots + \xi^{1} [C_{k}]$. Let 
$({\bf m},{\bf v})$ be an oriented edge corresponding with a non -- oriented
edge on which the cycle $\xi^{1} $ takes the value $1$. Then the vertex
${\bf m} + {\bf v}$ is incident to two or four non -- oriented edges on
which the cycle $\xi^{1}$ takes the value $1$. Let us choose an oriented edge
$({\bf m} + {\bf v},{\bf v}_{1})$, ${\bf v}_{1} \neq - {\bf v}$,
corresponding with a non -- oriented edge on which the cycle $\xi^{1} $
takes the value $1$. By repeating this process we obtain the prime closed
path $C_{1}$ such that the support of the cycle $\xi^{1} [C_{1}]$
is contained in the support of the cycle $\xi^{1} $. Hence there exists
a cycle $\eta^{1} \in Z_{1}(P(G),{\bf Z}_{2}^{add})$ such that
$\xi^{1} = \xi^{1} [C_{1}] + \eta^{1} $ and the supports of the cycles
$\xi^{1} [C_{1}]$ and $\eta^{1} $ do not intersect. By applying the above
procedure for a cycle $\eta^{1} $ we construct a prime closed path $C_{2}$.
By repeating this process we obtain  the disjoint prime non -- oriented
cycles $[[C_{1}]]$,...,$[[C_{k}]]$ such that
$\xi^{1} = \xi^{1} [C_{1}] + \cdots + \xi^{1} [C_{k}]$.

For a cycle $\xi^{1} \in B_{1}(P(G),{\bf Z}_{2}^{add})$ we can construct
the disjoint prime non -- oriented cycles $[[C_{1}]]$,...,$[[C_{k}]]$ 
such that $\xi^{1} = \xi^{1} [C_{1}] + \cdots + \xi^{1} [C_{k}]$ and
the prime closed paths satisfy the above condition. If 
$\xi^{1} \in B_{1}(P(G),{\bf Z}_{2}^{add})$, then 
$\xi^{1} = \partial \xi^{2} $ where a cochain
$\xi^{2} \in C^{2}(P(G),{\bf Z}_{2}^{add})$. The support of the cochain
$\xi^{2} $ consists of the squares on which the cochain $\xi^{2} $ takes
the value $1$. Two squares $s_{i}^{2}$ and $s_{j}^{2}$ belong to one
connected component of the support of the cochain $\xi^{2} $ if
there exist the squares $s_{i_{1}}^{2}$,...,$s_{i_{k}}^{2}$ from
the support of the cochain $\xi^{2} $ such that the boundaries of the
squares $s_{i}^{2}$ and $s_{i_{1}}^{2}$, $s_{i_{l}}^{2}$ and 
$s_{i_{l + 1}}^{2}$, $l = 1,...,k - 1$, $s_{i_{k}}^{2}$ and $s_{j}^{2}$
contain the common non -- oriented edges. The boundaries of the squares
from the different connected components of the support of the cochain
$\xi^{2} $ may contain the common vertices only. Thus
$\xi^{1} = \partial \xi_{1}^{2} + \cdots + \partial \xi_{k}^{2} $ where
the support of the cochain $\xi_{i}^{2} $ has the only connected component
and for $i \neq j$ the supports of the cochains $\xi_{i}^{2} $ and
$\xi_{j}^{2} $ do not intersect. The support of the cochain 
$\partial \xi_{i}^{2} $ corresponds with the prime non -- oriented cycle
$[[C_{i}]]$ on the graph $G$. By the construction the prime non -- oriented
cycles $[[C_{1}]]$,...,$[[C_{k}]]$ are disjoint. Moreover, if four
different oriented edges from the prime closed paths $C_{1}$,...,$C_{k}$ are
incident to one vertex, then the prime closed paths $C_{1}$,...,$C_{k}$ go
subsequently through such oriented edges from these four oriented
edges that have the directions orthogonal to each other.

Let a prime non -- oriented cycle $[[C]]$ be the boundary of a connected
set of the squares. We remove one square from this set, so that the new
set is also connected. Let a prime non -- oriented cycle $[[C^{\prime }]]$
be the boundary of this new connected set of the squares. Let $C$ and
$C^{\prime }$ be the representatives of the prime non -- oriented cycles
$[[C]]$ and $[[C^{\prime }]]$. By using the definitions (\ref{2.17}), 
(\ref{4.5}) it is possible to prove that $\rho (C) = \rho (C^{\prime })$.
By repeating this process we obtain $\rho (C^{\prime \prime}) = \rho (C)$
where $C^{\prime \prime}$ is a representative of the boundary of one 
square. It is easy to calculate that 
$\rho (C^{\prime \prime}) = \rho (C) = - 1$.

The previous arguments imply that the sum of the terms (\ref{5.36})
corresponding with a cycle $\xi^{1} \in B_{1}(P(G),{\bf Z}_{2}^{add})$
is equal to a unique term (\ref{5.36}) where $\rho (C_{i}) = - 1$,
$i = 1,...,k$. For a cycle $\xi^{1} \notin B_{1}(P(G),{\bf Z}_{2}^{add})$
we obtain a unique term (\ref{5.36}) where $\rho (C_{i}) = \pm 1$,
$i = 1,...,k$. The interaction energy $E({\bf e})$ is non -- negative.
Therefore we get the estimate
\begin{equation}
\label{5.39}
Z_{r,G} -
\exp \{ - 1/2\sum_{C \in RC(G)} |C|^{- 1}{\bf u}^{C}\rho (C)\} \leq
2\sum_{{\xi^{1} \in Z_{1}(P(G),Z_{2}^{add}),} \atop
{\xi^{1} \notin B_{1}(P(G),Z_{2}^{add})}} {\bf u}^{\xi^{1} }.
\end{equation}

Let us evaluate the right hand side of the inequality (\ref{5.39}).
We have proved that any cycle $\xi^{1} \in Z_{1}(P(G),{\bf Z}_{2}^{add})$
has the form $\xi^{1} = \xi^{1} [C_{1}] + \cdots + \xi^{1} [C_{k}]$
where the prime non -- oriented cycles $[[C_{1}]]$,...,$[[C_{k}]]$ are
disjoint. If $\xi^{1} \notin B_{1}(P(G),{\bf Z}_{2}^{add})$, then at 
least for one prime non -- oriented cycle $[[C_{i}]]$ the cochain
$\xi^{1} [C_{i}] \notin B_{1}(P(G),{\bf Z}_{2}^{add})$. Hence any cycle
$\xi^{1} \notin B_{1}(P(G),{\bf Z}_{2}^{add})$ has the form
$\xi^{1} = \xi^{1} [C] + \eta^{1} $ where the prime non -- oriented 
cycle $[[C]]$ corresponds with the cycle 
$\xi^{1} [C] \notin B_{1}(P(G),{\bf Z}_{2}^{add})$ and the supports of the
path $C$ and of the cycle $\eta^{1} \in Z_{1}(P(G),{\bf Z}_{2}^{add})$
do not intersect. By using this decomposition we obtain the estimate
\begin{equation}
\label{5.40}
\sum_{{\xi^{1} \in Z_{1}(P(G),Z_{2}^{add}),} \atop
{\xi^{1} \notin B_{1}(P(G),Z_{2}^{add})}} {\bf u}^{\xi^{1} } \leq
Z_{r,G}\sum_{{[[C]]: \, \, prime,} \atop 
{\xi^{1} [C] \notin B_{1}(P(G),Z_{2}^{add})}} {\bf u}^{C}.
\end{equation}
The total number of all reduced closed paths of the length $l$ with
the fixed initial vertex on the lattice 
$\tilde{G} (M_{1}^{\prime },M_{2}^{\prime };M_{1},M_{2})$ is less than
$4\cdot 3^{l - 1}$. The total number of vertices of the lattice
$\tilde{G} (M_{1}^{\prime },M_{2}^{\prime };M_{1},M_{2})$ is equal to
$(M_{1} - M_{1}^{\prime })(M_{2} - M_{2}^{\prime })$. Hence the total
number of all reduced closed paths of the length $l$ on the lattice
$\tilde{G} (M_{1}^{\prime },M_{2}^{\prime };M_{1},M_{2})$ is less than
$(M_{1} - M_{1}^{\prime })(M_{2} - M_{2}^{\prime })4\cdot 3^{l - 1}$.
Let for a prime closed path $C$ on the lattice
$\tilde{G} (M_{1}^{\prime },M_{2}^{\prime };M_{1},M_{2})$ the cochain
$\xi^{1} [C] \notin B_{1}(P(G),{\bf Z}_{2}^{add})$. Hence the length
$|C|$ of a path $C$ is more than 
$M = \min (M_{1} - M_{1}^{\prime }, M_{2} - M_{2}^{\prime })$. It
implies the following estimate
\begin{eqnarray}
\label{5.41}
\sum_{{[[C]]: \, \, prime,} \atop 
{\xi^{1} [C] \notin B_{1}(P(G),Z_{2}^{add})}} {\bf u}^{C} \leq
\sum_{C \in RC(\tilde{G} ), \, \, |C| \geq M} ||u||^{|C|} \leq \nonumber \\
4/3(M_{1} - M_{1}^{\prime })(M_{2} - M_{2}^{\prime })
\sum_{l = M}^{\infty } (3||u||)^{l} \leq \nonumber \\
4/3\bigl( \prod_{s = 1}^{2} (M_{s} - M_{s}^{\prime })\bigr)
(1 - 3||u||)^{- 1}\sum_{s = 1}^{2} (3||u||)^{M_{s} - M_{s}^{\prime }} 
\end{eqnarray}
where $||u||$ denotes the left hand side of the inequality (\ref{2.30}).
The inequalities (\ref{5.38}) -- (\ref{5.41}) imply the inequalities
(\ref{5.25}). The theorem is proved.

If a graph $G$ is embedded in a rectangular lattice on the plane, then
$Z_{1}(P(G),{\bf Z}_{2}^{add}) = B_{1}(P(G),{\bf Z}_{2}^{add})$.
Therefore the inequality (\ref{5.39}) becomes the equality (\ref{4.7}).
Thus we obtain a new proof of the equality (\ref{4.7}) independent of the 
papers \cite{5}, \cite{6}, \cite{12}, \cite{13}.

It follows from the relations (\ref{4.1}) -- (\ref{4.4}) for the
homogeneous Ising model on the torus that
\begin{equation}
\label{5.42}
Z_{\tilde{G} (M_{1}^{\prime },M_{2}^{\prime };M_{1},M_{2})} =
Z_{\tilde{G} (0,0;M_{1} - M_{1}^{\prime },M_{2} - M_{2}^{\prime })}.
\end{equation}
Let us denote $E_{1}$($E_{2}$) the interaction energy $E({\bf e})$ for
horizontally (vertically) directed edges ${\bf e}$ of the lattice
$\tilde{G} (0,0;M_{1},M_{2})$.

\noindent {\bf Theorem 5.4.} {\it Let for the non -- negative interaction
energy of the homogeneous Ising model on a rectangular lattice on the
torus the estimate} (\ref{4.6}) {\it be valid. Then for the partition
function} (\ref{4.1}) {\it of the homogeneous Ising model on the
rectangular lattice} $\tilde{G} (0,0;M_{1},M_{2})$ {\it on the torus} 
\begin{eqnarray}
\label{5.43}
\lim_{{M_{1},M_{2} \rightarrow \infty,}\atop 
{M_{1}(M_{2})^{- 1} + M_{2}(M_{1})^{- 1} \leq const}}
(M_{1}M_{2})^{- 1}\ln Z_{\tilde{G} (0,0;M_{1},M_{2})} = \nonumber \\
\ln (2\cosh \beta E_{1}\cosh \beta E_{2}) + 
1/2 (2\pi )^{- 2} \int_{0}^{2\pi } d\theta_{1} \int_{0}^{2\pi } d\theta_{2}
\nonumber \\
\ln [(1 + z_{1}^{2})(1 + z_{2}^{2}) - 2z_{1}(1 - z_{2}^{2})\cos \theta_{1} -
2z_{2}(1 - z_{1}^{2})\cos \theta_{2} ]
\end{eqnarray}
{\it where the variables} $z_{i} = \tanh \beta E_{i}$, $i = 1,2$.

\noindent {\it Proof.} The equality (\ref{2.36}) and the inequalities
(\ref{5.25}) imply
\begin{eqnarray}
\label{5.44}
\ln [1 - 8/3 (M_{1} M_{2})
(1 - 3||u||)^{- 1}\sum_{s = 1}^{2} (3||u||)^{M_{s}}]
\leq \nonumber \\
1/2\ln [\det (I - T({\bf u},\rho ))] - \ln Z_{\tilde{G} (0,0;M_{1},M_{2})}
\leq 0
\end{eqnarray}
where the matrix $T({\bf u},\rho )$ is given by the equalities (\ref{2.32})
and (\ref{4.5}) for the lattice 

\noindent $\tilde{G} (0,0;M_{1},M_{2})$. Now the
equality (\ref{5.43}) follows from the inequalities (\ref{5.44}) and
from the equalities (\ref{4.1}) and (\ref{4.26}). The theorem is proved.

The equality (\ref{5.43}) is proved in the paper \cite{11} for the
arbitrary interaction energies $E_{i}$, $i = 1,2$.

The relation (\ref{4.27}) is valid also for the graph 
$\tilde{G} = \tilde{G} (M_{1}^{\prime},M_{2}^{\prime};M_{1},M_{2})$.

\noindent {\bf Theorem 5.5.} {\it Let for Ising model on the rectangular
lattice} $\tilde{G} (M_{1}^{\prime},M_{2}^{\prime};M_{1},M_{2})$
{\it on the torus the estimate} (\ref{4.6}) {\it be valid and let the
interaction energy} $E({\bf e})$ {\it be non -- negative. Let a cochain}
$\chi \in C^{0}(P({\bf Z}^{\times 2}),{\bf Z}_{2}^{add})$ {\it be equal to}
$1$ {\it on the finite number of the vertices. Then for the correlation 
function} (\ref{3.14}) {\it of the two dimentional Ising model with
periodic boundary conditions}
\begin{eqnarray}
\label{5.45}
\lim_{{M_{s} - M_{s}^{\prime} \rightarrow \infty, \, \, s = 1,2,}
\atop {(M_{1} - M_{1}^{\prime})(M_{2} - M_{2}^{\prime})^{- 1} +
(M_{2} - M_{2}^{\prime})(M_{1} - M_{1}^{\prime})^{- 1} \leq const}}
W_{\tilde{G} (M_{1}^{\prime},M_{2}^{\prime};M_{1},M_{2})} (\chi ) =
\nonumber \\
\sum_{{\xi^{1} \in C^{1}(P(G),Z_{2}^{add}),\, \partial \xi^{1} = \chi ,}
\atop {\chi - connected \, ||\xi^{1} ||}} {\bf u}^{\xi^{1} }
\exp \{ 1/2\sum_{{C \in RC(Z^{\times 2}),}
\atop {||C||\cap i(||\xi^{1} ||) \neq \emptyset }}
|C|^{- 1} {\bf u}^{C}\rho (C) \}
\end{eqnarray}
{\it where the number} ${\bf u}^{\xi^{1} }$ {\it is defined by the 
relations} (\ref{4.3}), (\ref{4.4}), {\it the number} ${\bf u}^{C}$
{\it is defined by the relations} (\ref{2.18}), (\ref{4.4}) {\it and the
number} $\rho (C)$ {\it is defined by the relations} (\ref{2.17}), 
(\ref{4.5}).

\noindent {\it Proof.} Let the estimate (\ref{4.6}) be valid. Let the
interaction energy be non -- negative. Let a cochain
$\xi^{1} \in C^{1}(P(\tilde{G} ),{\bf Z}_{2}^{add})$ satisfy the condition
$\partial \xi^{1} = \chi $ and let its support $||\xi^{1} ||$ be
$\chi $ -- connected. Let $i(||\xi^{1} ||)$ be the set of all non -- oriented
edges incident to the vertices incident to the edges of the support
$||\xi^{1} ||$. The inequalities (\ref{5.25}) for the graphs $\tilde{G} $
and $\tilde{G} \setminus i(||\xi^{1} ||)$ imply
\begin{eqnarray}
\label{5.46}
1 - 8/3 \bigl( \prod_{s = 1}^{2} (M_{s} - M_{s}^{\prime })\bigr)
(1 - 3||u||)^{- 1}\sum_{s = 1}^{2} (3||u||)^{M_{s} - M_{s}^{\prime }} \leq
\nonumber \\
Z_{r,\tilde{G} }(Z_{r,\tilde{G} \setminus i(||\xi^{1} ||)})^{- 1}
\exp \{ 1/2\sum_{{C \in RC(\tilde{G} ),}
\atop {||C||\cap i(||\xi^{1} ||) \neq \emptyset }}
|C|^{- 1} {\bf u}^{C}\rho (C) \} \leq \nonumber \\
(1 - 8/3 \bigl( \prod_{s = 1}^{2} (M_{s} - M_{s}^{\prime })\bigr)
(1 - 3||u||)^{- 1}\sum_{s = 1}^{2} (3||u||)^{M_{s} - M_{s}^{\prime }})^{- 1}.
\end{eqnarray} 
The interaction energy $E({\bf e})$ is non -- negative. Then the 
definition (\ref{4.2}) implies the following estimate
\begin{equation}
\label{5.47}
(Z_{r,\tilde{G} })^{- 1}Z_{r,\tilde{G} \setminus i(||\xi^{1} ||)} \leq 1.
\end{equation}
It follows from the estimates (\ref{4.6}) and (\ref{5.47}) that for
$M_{s} - M_{s}^{\prime } \rightarrow \infty $, $s = 1,2$, the non -- zero
contributions give only those terms of the sum (\ref{4.27}) for the graph
$\tilde{G} $ which correspond to the cochains
$\xi^{1} \in C^{1}(P({\bf Z}^{\times 2}),{\bf Z}_{2}^{add})$ with the 
finite supports $||\xi^{1} ||$. Let a cochain
$\xi^{1} \in C^{1}(P({\bf Z}^{\times 2}),{\bf Z}_{2}^{add})$ satisfy the
condition $\partial \xi^{1} = \chi $ and let it have the finite
$\chi $ -- connected support $||\xi^{1} ||$. For sufficiently large
$M_{s} - M_{s}^{\prime}$, $s = 1,2$, the set $i(||\xi^{1} ||) \subset 
\tilde{G} (M_{1}^{\prime},M_{2}^{\prime};M_{1},M_{2})$. Let us prove that
\begin{eqnarray}
\label{5.48}
\lim_{{M_{s} - M_{s}^{\prime} \rightarrow \infty, \, \, s = 1,2,}
\atop {(M_{1} - M_{1}^{\prime})(M_{2} - M_{2}^{\prime})^{- 1} +
(M_{2} - M_{2}^{\prime})(M_{1} - M_{1}^{\prime})^{- 1} \leq const}}
(Z_{r,\tilde{G} })^{- 1}Z_{r,\tilde{G} \setminus i(||\xi^{1} ||)} =
\nonumber \\
\exp \{ 1/2\sum_{{C \in RC(Z^{\times 2}),}
\atop {||C||\cap i(||\xi^{1} ||) \neq \emptyset }}
|C|^{- 1} {\bf u}^{C}\rho (C) \}.
\end{eqnarray}
Indeed, for $M_{s} - M_{s}^{\prime} \rightarrow \infty $, $s = 1,2$, and
$(M_{1} - M_{1}^{\prime})(M_{2} - M_{2}^{\prime})^{- 1} +
(M_{2} - M_{2}^{\prime})(M_{1} - M_{1}^{\prime})^{- 1} \leq const$ 
the left and the right hand sides of the inequalities (\ref{5.46}) tend to 
$1$. Since the set $i(||\xi^{1} ||)$ is finite, the last multiplier in the 
central part of the inequalities (\ref{5.46}) tends to the right hand side 
of the equality (\ref{5.48}). The series (\ref{5.45}) coincides with the 
series (\ref{4.28}). It was proved in Theorem 4.3 that the series (\ref{4.28})
is absolutely convergent if the estimate (\ref{4.6}) is fulfilled and
the interaction energy is non -- negative. Thus the sum (\ref{4.27})
for the graph $\tilde{G} (M_{1}^{\prime},M_{2}^{\prime};M_{1},M_{2})$
converges to the series (\ref{5.45}). The theorem is proved.

The correlation functions (\ref{5.45}) and (\ref{4.28}) coincide.

\end{document}